\documentclass[12pt,preprint]{aastex}

\shorttitle{Dwarf galaxies in NGC 2784 Group}
\shortauthors{Park et al.}

\begin{document}

\title{Dwarf Galaxy Discoveries from the KMTNet Supernova Program{*} I. The NGC 2784 Galaxy Group}

\author{ Hong Soo Park\altaffilmark{1,2}, 
Dae-Sik Moon\altaffilmark{3},
Dennis Zaritsky\altaffilmark{4}, 
Mina Pak\altaffilmark{1,2},
Jae-Joon Lee\altaffilmark{1},
Sang Chul Kim\altaffilmark{1,2}, 
Dong-Jin Kim\altaffilmark{1}, and 
Sang-Mok Cha\altaffilmark{1}
}

\email{hspark@kasi.re.kr
}

\altaffiltext{1}{Korea Astronomy and Space Science Institute, 776, Daedeokdae-ro, Yuseong-gu, Daejeon 34055, Korea}
\altaffiltext{2}{Korea University of Science and Technology (UST), Daejeon 34113, Korea}
\altaffiltext{3}{Dept. of Astronomy and Astrophysics, University of Toronto, 50 St. George Street, Toronto, ON M5S 3H4, Canada}
\altaffiltext{4}{Steward Observatory, University of Arizona, 933 North Cherry Avenue, Tucson, AZ 85719, USA}
\altaffiltext{*}{Based on data collected at KMTNet Telescopes}
 
\begin{abstract}

We present $BVI$ surface photometry of 31 dwarf galaxy candidates discovered 
in a deep image stack from the KMTNet Supernova Program of $\sim$ 30 square degrees centered on the nearby NGC 2784 galaxy group.
Our final images have a 3$\sigma$ surface brightness detection limit of $\mu_V\approx 28.5$ mag arcsec$^{-2}$. The  
faintest central surface brightness that we measure is
$\mu_{0,V} = 26.1$  mag arcsec$^{-2}$. If these candidates are at the distance of 
NGC 2784, then they have absolute magnitudes greater than $M_V = -9.5$ mag
and effective radii larger than 170 pc.
Their radial number density decreases exponentially with
distance from the center of NGC 2784 until it flattens
beyond a radius of 0.5 Mpc. We interpret the baseline density level to represent the background contamination and so
estimate that 22 of the 31 new candidates are dwarf members of the group.
The candidate's average color,  $\langle (B-V)_0\rangle\approx 0.7$, 
and S{\'e}rsic structural parameters are consistent with those parameters for the dwarf populations of
other groups. We find that the central population of dwarfs is redder and brighter
than the rest of the population. The measured faint end slope of the luminosity function,  $\alpha\approx-1.33$,
is steeper than that of the Local Group but consistent with published results for other groups. Such comparisons are complicated
by systematic differences among different studies, but will be simpler when the KMTNet survey, which will provide homogenous data
for 15 to 20 groups, is completed.

\end{abstract}


\section{Introduction}

Dwarf galaxies in nearby  ($D\lesssim$ 10 Mpc) galaxy groups are 
probes of the evolution of low mass galaxies in low density environments.
Their properties, such as the luminosity function or radial distribution within a group,
are often 
used as tests of conventional hierarchical cosmology models.
For example, 
one challenge to the simple standard paradigm is that the number of dwarf galaxies predicted in a $\Lambda$CDM model is
  much larger than that observed near the Milky Way \citep{kly99, moo99}.
 Various solutions to this problem have been suggested, such as
suppression
of galaxy formation by photoionization \citep{ben02, coo10} 
and a steeply declining relation between stellar mass and halo mass \citep{beh10},
and
the problem has lessened with the subsequent discovery of 
numerous faint Milky Way dwarf galaxies in 
surveys like the Sloan Digital Sky Survey \citep{mcc12} and the Dark Energy Survey \citep{bec15,drl15}.
Nevertheless, the apparent conflict helped focus theoretical work on aspects of
  the quenching of star formation in low-mass group galaxies 
  \citep{wet12, pre12} and observational work on addressing deficits in the original data. 

Several recent searches for 
  new dwarf galaxies
  in nearby groups 
  have exploited CCD cameras with wide fields of view ($>$ 1 $\square^\circ$).
 \citet{chi09} reported 22 new dwarf galaxies found in an area of $\sim$ 65 $\square^\circ$ around the M81 group
using CFHT/MegaCam. 
 \citet{mul15} found 16 new dwarf galaxies 
 in an area of $\sim$ 60 $\square^\circ$ around M83 
 using Dark Energy Camera (DECam) at the CTIO 4-m telescope.
 The same group also reported $\sim$ 40 new dwarf galaxy candidates in the Centaurus group region \citep{mul17}.
Other studies have explored the 
  NGC 5128 group \citep{crn16}, the 
  M 106 group \citep{kim11}, the NGC 1023 group \citep{tre09}, and the 
  M 101 group \citep{mer14}.
These studies all discovered new dwarf galaxies, reaching as faint as $M_V \sim -10$ mag. 
An aim of these studies is to establish the general properties of dwarf galaxies in groups and identify any variations that 
depend on group properties.
However, despite all of this recent progress it is difficult to construct a coherent picture because 
comparing among different studies is undermined by systematic differences in data quality, selection criteria, and methodology.
So motivated, we are carrying out a search for 
dwarf galaxies around other nearby galaxy groups using a single, large survey. We present our first results
here.

The NGC 2784 group contains the lenticular-type NGC 2784 galaxy
 and at least ten other galaxies \citep{fou92, gar93, kar13}. 
The distance measurement, 9.8 Mpc, comes from applying the surface brightness fluctuation method \citep{ton01}.
This galaxy group is
among the fields targeted by the KMTNet (Korean Microlensing Telescope Network) Supernova Program \cite[KSP;][]{moo16}, 
which is an international supernovae search with associated follow-up photometry using three 1.6-m telescopes.
 The field of view of these telescopes is extremely wide (4 $\square^\circ$ per exposure)
 and we utilize the total exposure obtained over several hundreds of individual images 
 obtained for the supernova and optical transient \citep{he16} search.
  
This paper is organized as follows. In
  Section 2 we describe the observations and data reduction.
  In Section 3, we discuss the search technique and the surface photometry of the dwarf galaxy candidates.
  In Section 4, we present the properties of the NGC 2784 group and discuss our results in comparison to those published for other galaxy groups.

\section{Observations and Data Reduction}\label{data}
 
\begin{figure}
\epsscale{1.0}
\plotone{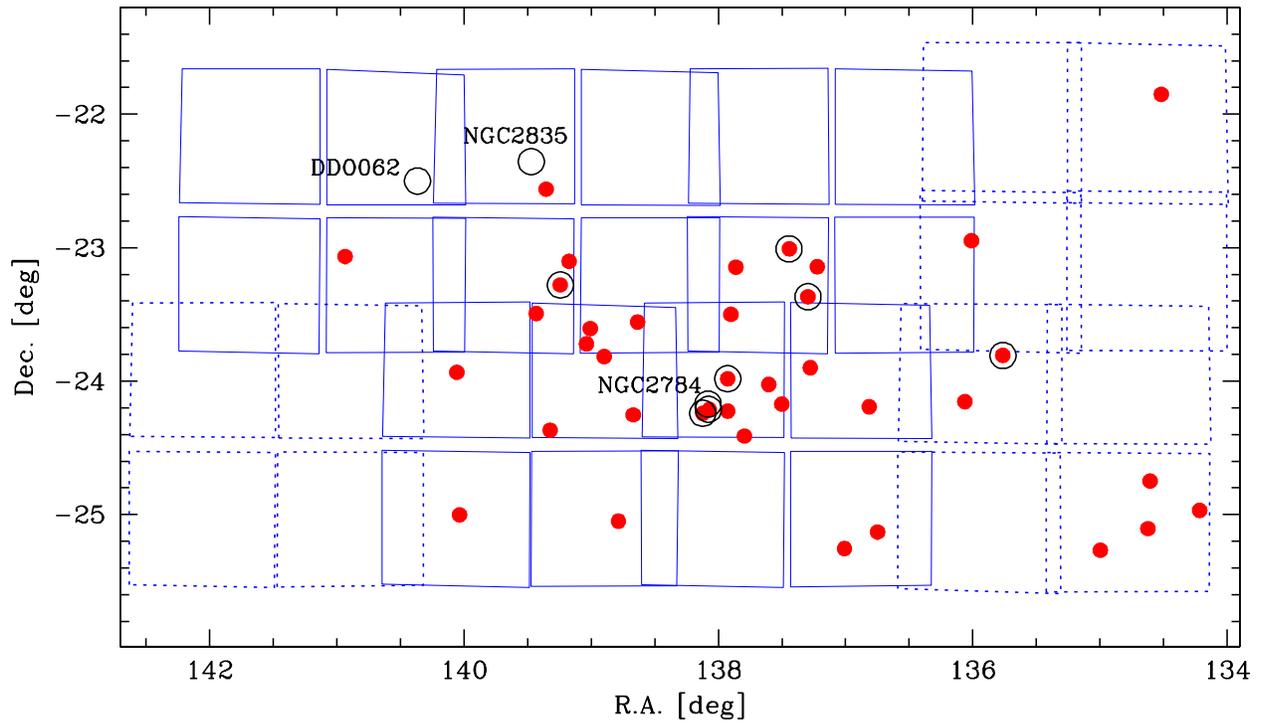}
\caption{ 
Finding chart for the NGC 2784 field.
The filled circles are dwarf galaxy candidates of the NGC 2784 field.
The open circles indicate nearby galaxies within 11 Mpc \citep{kar13} and
we labeled three  of the brightest galaxies in this field. 
The large solid and dotted 
squares indicate observed individual fields with approximately $1^\circ \times 1^\circ$ area
for deep and shallow stack images, respectively.
\label{fig-radec}}
\end{figure}

We carried out $BVI$ observations of eight fields around NGC 2784 using the KMTNet 
facility \citep{kim16} between February and July of 2015.
KMTNet is a network of three new wide-field 1.6-m telescopes, located at 
the Cerro Tololo Inter-American Observatory (CTIO, Chile), 
the South African Astronomical Observatory (SAAO, South Africa), 
and Siding Spring Observatory (SSO, Australia). 
Each telescope is equipped with a wide-field CCD camera covering 
4 $\square^\circ$ (2$^\circ \times 2^\circ$) with 0.4$\arcsec$ pixel$^{-1}$ sampling.
The detector is a mosaic of four 9K $\times$ 9K E2V chips
with a small gap between chips,
and each chip corresponds to $1^\circ \times 1^\circ$ field.
We used $BVI$ filters and the exposure time for an individual frame was fixed to be 60 sec 
for any of the filters. In 
Figure \ref{fig-radec} we overlay the positions of the dwarf galaxy candidates discovered in this work (see Section 3) 
on top of the eight observed fields.
We observed five of the eight fields several hundred times through each filter and the remaining
three fields only tens of times. 

After pre-processing the raw images, which includes crosstalk correction, overscan subtraction, 
and flat fielding, we do the astrometric 
calibration using the $SCAMP$ routine \citep{ber06} 
with the GSC-2.3 reference catalog
for each of the four 1$^\circ \times 1^\circ$ chips.
We then apply the $SWarp$ routine  \citep{ber02} to create deep, stacked images. 
At that time, to subtract the sky, we use $SWarp$'s background subtraction and a mesh size with 512 pixels.
We only use images with seeing better than 2 arcsec, resulting in an
average seeing for our final stacked images that is better than 1.5 arcsec in all filters.
Tables \ref{tab-obslog1} and \ref{tab-obslog2} list the details of the fields and observations.

\begin{figure}
\epsscale{0.9}
\plotone{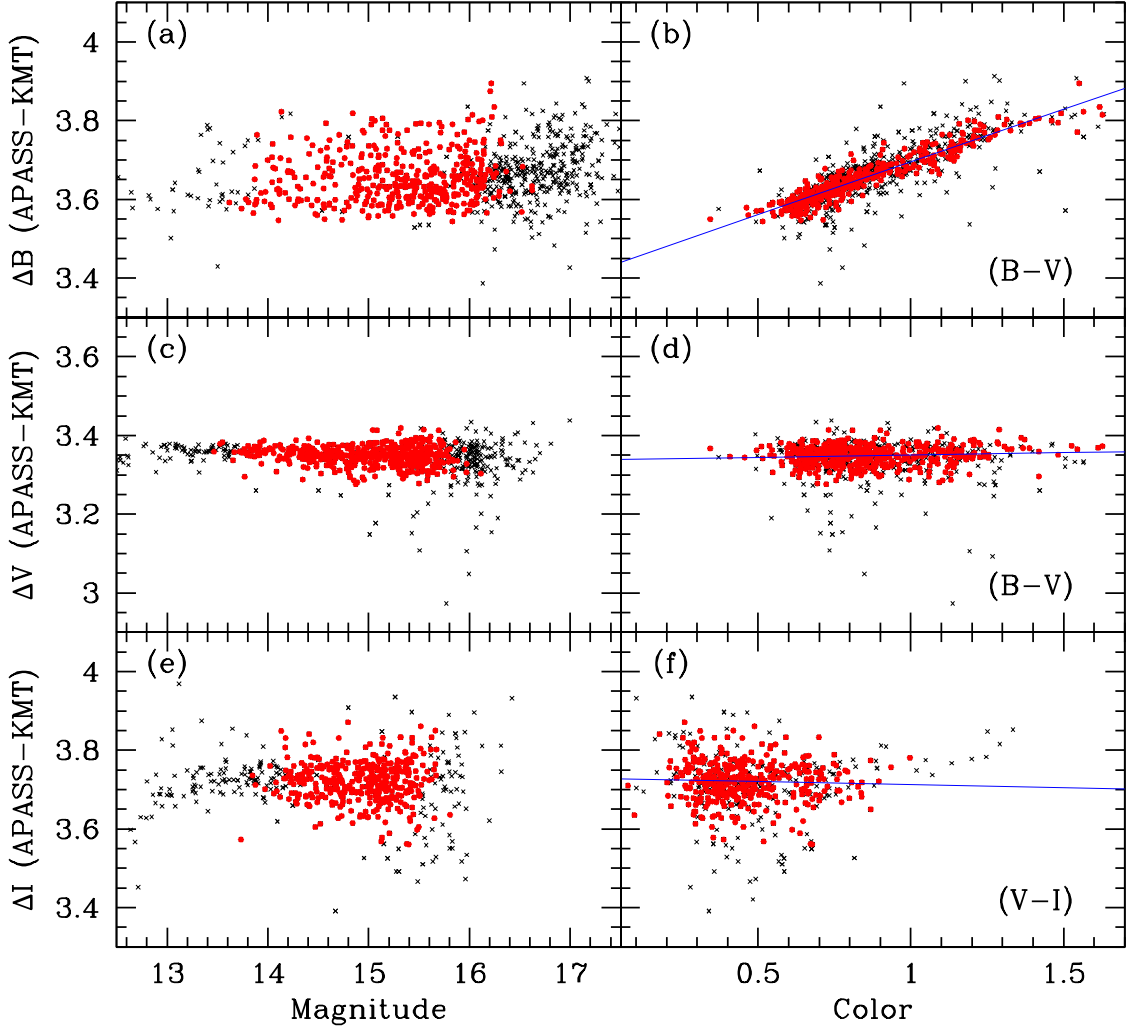}
\caption{
Magnitude difference between APASS standard magnitude and KMTNet instrumental magnitude
as a function of magnitude (left panels) and color (right panels).
The data with the red filled circles and black crosses are the point sources matched between APASS catalog and KMTNet photometry catalog.  
The solid lines represent the linear least-square fits to the red filled circles, 
which indicate the data within $3\sigma$ from the fits with too bright or too fainter stars are excluded.
\label{fig-stdcal}}
\end{figure}

We derive transformation equations      
 between instrumental magnitudes ($b, v, i$) and standard Johnson-Cousins magnitudes ($B, V, I$) 
 using the APASS (AAVSO Photometric All-Sky Survey) catalog \citep{apass}.
 We use the equations as follows:
\begin{equation}\label{eq-std}
  B = b + c(B-V) + zero_B,~
  V = v + zero_V,~  
  I = i + zero_I.
\end{equation}
\noindent
  Here, $c$ is the color coefficient and $zero$ is the zero point magnitude for each filter.
  The $I$-band magnitude is referenced to the Sloan $i$-band magnitude
    because the APASS catalog includes only Sloan $i$-band magnitudes.
We derive transformation equations for each chip 
  by matching the magnitudes of bright point sources measured using $SExtractor$ \citep{ber96} 
  to the APASS catalog.     
In the case of the deep image stacks,
  $zero$ values are approximately 3.4, 3.4, and 3.7 magnitudes and 
  $rms$ values are about 0.02, 0.02, and 0.04 magnitudes for the $B, V, I$-bands, respectively.
In the case of the shallow image stacks,
  $zero$ values are approximately 4.0, 3.9, and 4.2 magnitudes and
  $rms$ values are about 0.03, 0.03, and 0.07 magnitudes for the $B, V, I$-bands, respectively.  
The $c$ color coefficient for both the deep or shallow image stacks is 0.27.
The number of point sources used in the calibration is about 300 per chip.  
Figure \ref{fig-stdcal} is an example (N2784-4-Q0 field including the NGC 2784 galaxy) of how differences in the pass bands between APASS and KMT propagate to color dependencies. 
The $B$-band only shows significant color dependency.
We note that the relatively larger scatter at the $I$-band is caused by the different filter system
between KMTNet and APASS.
We correct the measured magnitudes for foreground Galactic extinction of $A_B=0.761$, $A_V=0.575$, and $A_i=0.356$ $(A_I=0.316)$ \citep{sch11}\footnote{We pause here to clarify a variety of potentially confusing notation. First, we only apply extinction corrections to total magnitudes and colors. If such corrections have been applied, then we signify that with a subscript 0, as in $(B-V)_0$. If no subscript is included, then the corrections have not been applied. Second, for historical consistency we also use the subscript 0 to refer to central values, as in the central surface brightness, $\mu_0$. The subscript here does not imply that an extinction correction was applied. Third, we also use the subscript zero on a scaling radius, $r_0$, that is different from the effective
radius, $r_e$. The relationship between the scaling radius and $r_e$ depends on the S\'ersic index, $n$.}.


\section{Results}\label{result}

\subsection{Searching for Dwarf Galaxy Candidates}

Three of us (H. S. Park, M. Pak, and S. C. Kim) visually select dwarf galaxy candidates. We contend that
visual identification is more complete and less contaminated 
  than current automatic detection methods \citep{kim11, mul15}. 
Our selection criteria, applied to images displayed using the $zscale$ option in $DS9$, are that
  (1) there must be a diffuse stellar component that extends at least $\sim 10$ arcsec,
  (2) the object must not exhibit any distinct structure (e.g. such as spiral structure) except for a possible nuclear point source, and
  (3) the object must not have a steep surface brightness gradient between the center and the edge, such as that expected for 
      luminous background galaxies.
Our selection is mainly based on the $I$-band images, but we use 
  the $B$ and $V$-band images
  as well as the RGB color images made from the three bands
  to rule out artifacts.
HSP, MP, and SCK searched the images independently and then compared results. 
For the targets selected by only one or two classifiers, all three classifiers reviewed the object and reached a consensus on whether to retain
or reject the candidate.

\begin{figure}
\epsscale{1.0} 
\plotone{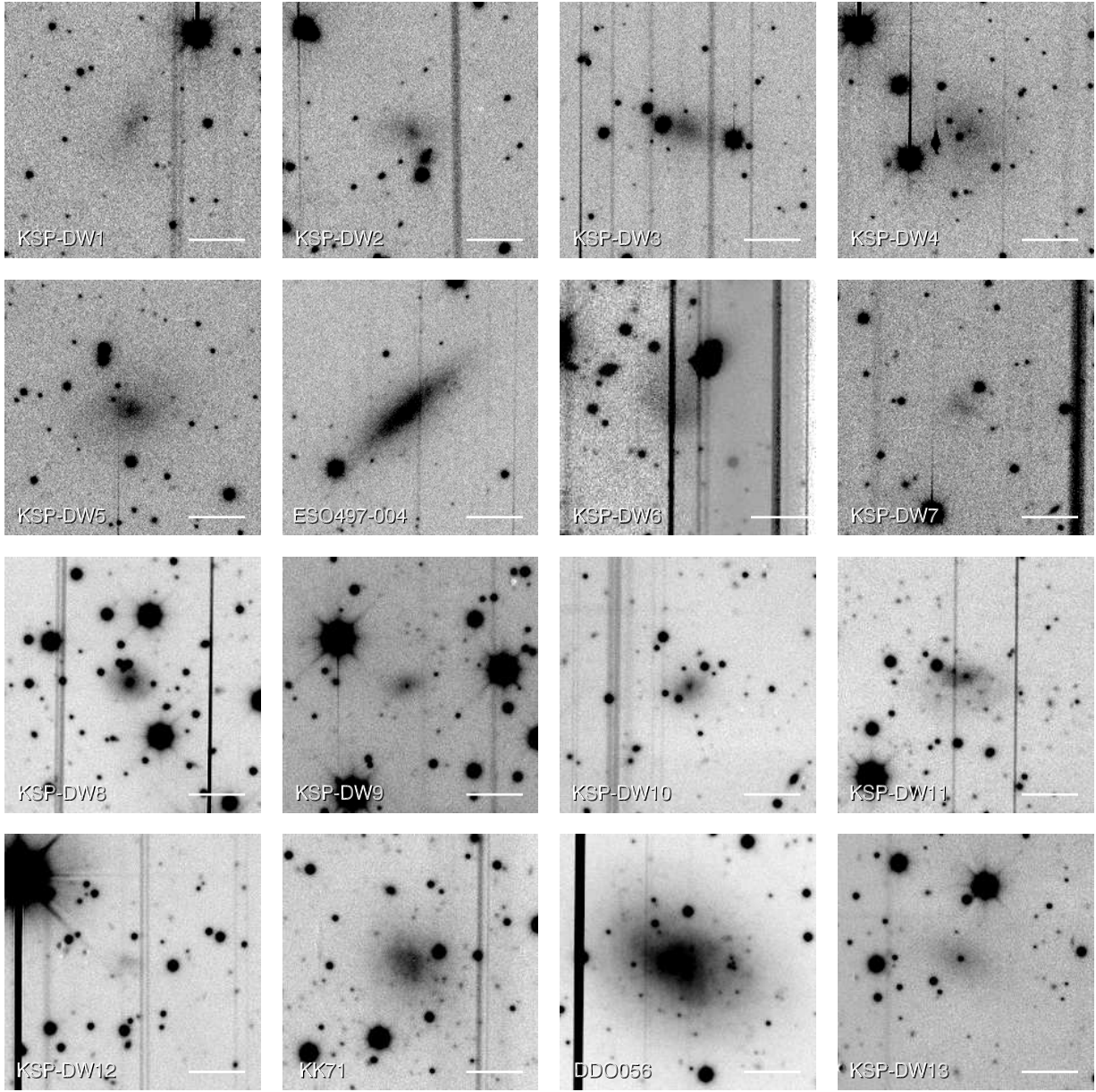}
\caption{$I$ band images of the dwarf galaxy candidates. 
The scale bar in each panel represents 20 arcsec, and  %
the field of view of most images is approximately 1.5 arcmin $\times$ 1.5 arcmin.
The last thumbnail figure represents an artificial dwarf
galaxy around the brightness with about 90\% completeness.
North is up, East is left.
\label{fig-graymapa}}
\end{figure}

We visually inspect each chip in each of the eight NGC 2784 fields, including both the shallow and deep image stacks.
We detect 38 objects that we classify as dwarf galaxy candidates and plot their 
positions in Figure \ref{fig-radec}. 
There are 7 previously identified dwarf candidates in the NGC 2784 fields \citep{kar13}. We recover all of them.
Three (ESO497-004, DDO 056, HIPASS J0916-23b) of these have been spectroscopically confirmed to be members \citep{doy05,fis81,kar08},
while the other four (KK71, KK72, NGC2784 dw01, KK73) await spectroscopic confirmation.
In addition, the group also contains three brighter galaxies (NGC 2784, NGC 2835, DDO 062) 
that have been spectroscopically confirmed as members.
We present $I-$band greyscale images of all the dwarf galaxy candidates in 
Figures \ref{fig-graymapa}.
Many of the images have evident vertical lines,
which are due to charge bleeds or blooming caused by saturation of the bright foreground stars on the CCD images. These artifacts do not affect our candidate detection 
because the sources are more extended nor do they affect
our surface photometry 
because we mask the artifacts (see Section \ref{photometry}).
The candidates exhibit a range of morphologies and sizes, and sometimes have an evident unresolved nuclear source.

We test our completeness and limiting brightness by inserting artificial sources in the images. We
adopt exponential surface brightness ($n=1$) profiles and central brightness and effective radii distributions that reproduce the effective radius -- magnitude trend of our NGC 2784 dwarf candidates (Figure \ref{fig-censurface} and Section \ref{cmdstructure}). 
We insert several hundred artificial galaxies 
in the  $I$-band image that includes NGC 2784
  using the {\it mkobjects} task in the IRAF{\it /artdata} package. We then reapply our
  visual inspection technique.
We find that for $m_I=18.9$ ($M_V=-11.0$ at the distance of NGC 2784) we are 90\% complete.
The last thumbnail image in Figure \ref{fig-graymapa} contains an artificial dwarf galaxy at this magnitude limit ($m_I=18.9$ mag and $r_{e}=17.5$ pixel).  
Several studies have shown that the surface brightness of dwarf galaxies can be described by a S{\'e}rsic profile and their S{\'e}rsic index distribution is peaked around $n\approx 0.7$ \citep{fer06, mun15}.
Thus, we also carry out the completeness test with $n=0.7$. 
We find that for $m_I=19.1$ ($M_V=-10.8$) we are 90\% complete.
We present detections down to $M_V \sim -10$, so we expect to be significantly incomplete at our limiting magnitude. In addition, our completeness simulations do 
not account for other possible populations. For example, we do not include ultra compact dwarf galaxies that we might
have considered to be stars or background galaxies. Therefore, the number 
38 is a lower limit on the number of the dwarf candidates in these fields.

\subsection{Surface Photometry}\label{photometry}

We measure the surface brightness profiles of the 38 dwarf candidates 
  in $B, V, I$ using the IRAF/{\it ELLIPSE} task. We fix
the center coordinates, ellipticities, and position angles, 
  which we estimate near the effective radius of each candidate.
We use linear semi-major axis steps of 1.0 arcsec  for the smaller candidates
and 1.5 arcsec for the larger ones.
The brightness of each isophote is measured from mean isophotal flux, and 
    its error is the 
    root-mean-square scatter of the isophotal flux values.
The instrumental brightnesses obtained from this surface photometry are transformed 
using the transformation equations (Equation \ref{eq-std} in Section \ref{data}).
  
  We present the surface brightness and color profiles in Figure \ref{fig-radimu0a}.
Most candidate have surface brightness profiles that are well described as exponential ($n\sim1$) and have constant color profiles. 
The 3$\sigma$ faintest surface brightnesses reached on our shallow and deep image stacks are
   $\sim \mu_I\approx 26.5$ ($\mu_V\sim 26.9$) mag arcsec$^{-2}$ and 
  $\mu_I\sim 28.0$ ($\mu_V\sim 28.5$) mag arcsec$^{-2}$, respectively.
In the color profile panels, 
the dot-dashed line represents the weighted mean value.
We adopt these mean values ($\langle\mu_{(V-I)}\rangle$ and $\langle\mu_{(B-V)}\rangle$)
   as the colors ($(B-V)$ and $(V-I)$) of the dwarf candidates.
 The total magnitudes are estimated from the  extrapolation of a fitted brightness profile, and
 $B$-band signal  is relatively much shallower due to same exposure time in each band ($B,V,I$). 
 Thus, in this study the mean value from  brightness profiles is more efficient for representing the color of each dwarf.
The color ranges are $ 0.6<(B-V)<1.2$ and $ 0.3<(V-I)<1.2 $.
    
We fit S{\'e}rsic functions
  ($\mu_0+1.0857(r/r_0)^{1/n}$) to the surface brightness profiles,
  where $\mu_0$, $r_0$, and $n$ are the
  central surface brightness, scale length, and S{\'e}rsic curvature index, respectively.
  We adopt
  the 1-sigma statistical errors computed from the non-linear least squares fitting procedure \citep{mar09}
  as the errors for the S{\'e}rsic parameters.
  We show these fits with solid curves in Figure \ref{fig-radimu0a}.
As a test, we compare the profiles of the
dwarf candidates that were observed in overlapping images (Figure \ref{fig-radec}). The resulting profile parameters are statistically consistent.
The faintest central surface brightnesses among the dwarf candidates are
    $\mu_{0,I}\sim 25.1$, $\mu_{0,V}\sim 26.1$, and $\mu_{0,B}\sim 26.9$ mag arcsec$^{-2}$.
  The S{\'e}rsic scale lengths range from  2$<r_{0,I}<$23 arcsec, and 
  the curvature indices from  $0.5<n_{I}<1.4$, with median value of $\sim$0.8 .
The total magnitudes of the dwarf candidates
  using these fits, following the method described in Section 3.1 of \citet{chi09}, result in the faintest
  $I_T$, $V_T$, and $B_T$ being $\sim$ 20.5, 20.6, and 21.7 mag, respectively.   
  The smallest $I$-band effective radius ($r_{e,I}$) is $\sim 3.5$ arcsec (corresponding to $\sim 170$ pc at the distance of NGC 2784).

\subsection{The NGC 2784 Dwarf Galaxy Catalog}\label{dwarfcat}

We present the photometric catalog %
  for the 38 dwarf candidates in  Table \ref{tab-cat}.
The first column contains our identification or the previously established galaxy name if available \citep{kar13}. The second and third columns
have the central  
R.A. (J2000) and Dec. (J2000).
The 4th column presents the $I$-band total magnitude derived from the S{\'e}rsic fit result.
The 5th and the 6th columns list the $(V-I)$ and $(B-V)$ colors, respectively.
Because the Galactic extinction values toward NGC 2784 are  depend on an adopted extinction map or calibration,
we present the colors without the extinction correction. 
The 7th,  8th, and  9th columns contain the central surface brightness ($\mu_{0,I}$), scale length ($r_{0,I}$), and curvature index ($n_{I}$) derived for the S{\'e}rsic fit, respectively.
The $I-$band $r_{e,I}$ as derived from our fits is in the 10th column.
We could not reliably measure  $(B-V)$ for KSP-DW6, and $(V-I)$ and $(B-V)$ for KSP-DW23
due to the low surface brightness of these galaxies. 

\subsection{Radial Number Density}

We present the 
number density of dwarf candidates in the NGC 2784 field
 as a function of the projected radius ($R$) 
 from the center of the stellar light of NGC 2784 in Figure \ref{fig-rnden}.
 Here, we carefully estimate the unit 'projected area' due to complex survey coverage (e.g. overlapped area, rectangular coverage on the whole) as seen in Figure \ref{fig-radec}, 
  and we adopt density errors assuming a Poisson distribution.
 The number density, 
 obtained from the visually identified dwarfs only in the region covered by deep stack images,
 decreases outwardly until the last bin
 ($3\lesssim R\lesssim 4$ degree) where it flattens.
 We interpret the baseline flat level as indicating that we have reached the background density of objects 
 that satisfy our selection criteria. It is possible that the profile of true dwarfs flattens somewhat
 at large radii, but conservatively we interpret this as the contamination level.
In contrast, when we examine the density profile 
using the region covered by shallow and deep stacks,
it rises at the outermost bin even though we expect the shallow stacks to be more incomplete.
However, most objects in this bin are concentrated in the south-west region and
located at $R\gtrsim 3$ degree from the NGC 2784. We suspect that these
candidates are contaminants.

When we fit the density distribution from only the deep stack results with an exponential profile plus a constant value, 
 we find that the surface density ($\Sigma$) is described by
 $\Sigma = e^{-1.62(\pm0.19) R + 2.52(\pm0.29)} + 0.24(\pm0.12)$.
The constant value (0.24 deg$^{-2}$) is the contamination level.
The density profile is also well described by a power-law with equal logarithmic interval (Figure \ref{fig-rnden}b). 
That fit has $\Sigma = 1.69(\pm0.47)R^{-1.27(\pm0.17)}+0.24$. 
This power index ($\alpha=-1.27\pm0.17$) is consistent with 
  the power index ($\alpha=-1.03\pm0.19$) obtained from an equal linear interval.
Integrating this exponential fit and comparing the number of dwarf candidates that are
radially clustered about NGC 2784 to those that are not suggests that 15\% (5)
of the 32 candidates within $R=2.7$ degree are not dwarfs associated with NGC 2784.
Additionally, 6 ($R>2.7$ degree) out of the 38 dwarf candidates may also be not associated with NGC 2784. 

\subsection{Comparison to Dwarf Galaxies in Other Systems}\label{cmdstructure}

In Figure \ref{fig-cmd} we present the color-magnitude diagram for the candidate dwarf population of NGC 2784.
The mean values of $(B-V)_0$ and $(V-I)_0$, which are corrected the Galactic extinction,  are
  $0.67\pm0.16$ and $0.45\pm0.20$, respectively. 
The $(B-V)_0$ mean color is similar to those for the M106  \citep[$\langle (B-V)_0\rangle\approx 0.73$;][]{kim11} 
  and the M83 groups \citep[$\langle (B-V)_0\rangle\approx 0.82$;][]{mul15}. 
For comparison, we overlay the values of dwarf galaxies in other groups 
(M83 from \citet{mul15}, M106 from \citet{kim11}) and 
the color-magnitude relations of the early-type galaxies in
clusters (Virgo from \citet{lis08}, and Ursa Major from \citet{pak14}).
We conclude that the color distributions are consistent and that the NGC 2784 dwarfs are mostly early-type galaxies.

We now compare the central surface brightnesses ($\mu_{0}$), effective radii ($r_{e}$), and S{\'e}rsic $n$ indices of the NGC 2784 dwarf candidates to 
those of dwarf galaxies in other groups.
For the central surface brightnesses and effective radii
  we use values from \citet{mul15} for the M83 group and from \citet{mcc12} for the Local Group.
  For $n$, 
  we use values from \citet{chi09} for the M81 group, 
  from \citet{mul15} for the M83 group, and from \citet{jer00} for the Local Group.
For our values, we calculate $\mu_{0,V}$ and $M_V$ using $\mu_{0,I}$ and $M_I$ in combination with the $(V-I)$ color.
We adopt $(V-I)=0.7$ for KSP-DW23.
The values for $r_{e}$ and $n$ are the values obtained from the $I$-band images.
We assume all of the candidates are at the distance of NGC 2784.
In Figure \ref{fig-censurface} we compare $\mu_{0,V}$, $r_{e}$, and $n$ 
  as a function $M_V$. 
The properties of the NGC 2784 dwarf candidates, with central surface brightness and effective radius increasing as the total brightness increases, agree well with the trends of the dwarfs in the M83 group and the Local Group.
These results support the contention that the candidates are mostly member galaxies of the NGC 2784 group rather than background galaxies.
Additionally, the median value of the S{\'e}rsic curvature index for the NGC 2784 dwarf candidates ($n\approx 0.8\pm 0.1$) is  similar to that of the dwarfs in the M83 group ($n\approx 1.1\pm 0.3$), the M81 group ($n\approx 0.6\pm 0.2$), and the Local Group ($n\approx 1.2\pm 0.3$).
Again, the properties of the NGC 2784 candidates are entirely consistent with those
  of dwarfs in other groups. 

\section{Discussion}\label{discuss}

\subsection{The NGC 2784 Galaxy Group}\label{discuss-n2784}

In Figure \ref{fig-rndenmag} we show the radial number density profiles of the bright ($M_V<-13$) and faint ($M_V\geq-13$) dwarf galaxy candidates in the NGC 2784 field.
We fit each profile with an exponential function for $R\lesssim 2.5^\circ$.
We find that the density profile of the brighter dwarfs 
($\ln(\Sigma)= -1.63 (\pm0.19) R + 1.78 (\pm0.24)$ $(rms=0.21)$)
is steeper than
that of the fainter dwarfs 
($\ln(\Sigma)= -0.68 (\pm0.24) R + 1.18 (\pm0.31)$ $(rms=0.26)$).
A complication in interpreting this result is that background contamination may be distorting the
fits. Although a magnitude dependent contamination is expected, our sample is too small for us to determine whether 
the background contamination is more significant for fainter candidates. If it is, however, it could be making the fainter population appear
more extended.  To test for such an effect,
we added the entire fitted background level (0.24 deg$^{-2}$, see Section 3.4) to the density profile of the brighter galaxies.
The new slope for this population is now $-1.28\pm0.19$, %
which is still steeper than that of the faint galaxies. 
We conclude, with some caution, that the brighter set of dwarf galaxies is 
more centrally concentrated.
This conclusion is further  supported by the results of a $Z$-test comparison of the radial distance distributions of the dwarf candidates within  $R = 2.5^\circ$.
$Z$-test is a statistical test to examine whether or not two samples are different population under assuming a normal distribution \citep{spr11}.
We find with 99.1\% confidence that the samples do not have the same radial distribution.
The difference in central concentration may reflect mass segregation, with more massive galaxies in a cluster or group being more centrally concentrated \citep{pre12, rob15}. Plausible physical 
reasons for that apparent concentration range from environmentally dependent formation histories, differences in dynamical friction, or 
mass-dependent tidal stripping.

Next, we check whether there are radial variations in the mean colors of the dwarf candidates.
In Figure \ref{fig-rcolor} we show that the dwarfs are redder when closer to the center. 
The trend is also visible for bright and faint dwarfs independently.
The radial fraction of red dwarf galaxies is higher in the center than in the outskirts when
we estimate the red dwarf fraction using the color-magnitude relation (equation (1)) by \citet{wei11}.
This observation is consistent with the results of simulations \citep{wet12, wet13} and other observations \citep{pop06, pre12, lee16}.
The standard interpretation of this effect is that 
the central galaxies have lower star formation rates. 

This result has also been reported in galaxy clusters such as Virgo and Coma \citep{wei11, smi12, lis13}. %
It is suggested that the galaxies in the central regions of massive clusters experienced earlier, more  efficient star formation quenching due to  stronger tidal forces and interactions with the intracluster medium, so the red galaxies are more centrally concentrated.   
The same environmental effect could be the reason for color distribution in the NGC 2784 group.
On the other hand, 
this reddening of galaxies toward the center might also be a metallicity effect. 
If, as discussed above, the central galaxies are more massive then they are likely to
also be more metal rich and therefore redder.
However, again this result might be influenced by differential background contamination.
In general, background contaminants tend to be blue galaxies, which are more common than red galaxies, and tend to be found at larger separations due to the greater amount of sky at larger separations. This effect could result in creating a color gradient in the sense of that observed.
To test for such an effect, we assume that  the 5 dwarf candidates
  (this number is based on the 15\% contamination rate with $R\lesssim 2.7$ degree (see Section 3.4)) 
  are all contaminants with bluish color and we remove them from our sample.
  The color trend remains.
 
\subsection{Cumulative Luminosity Functions of Galaxy Groups}

To investigate the luminosity function (LF) of galaxies in the NGC 2784 group and compare it to that 
of galaxies in other groups, 
we construct the cumulative LF of the NGC 2784 group.
We focus on the cumulative LF,  instead of the differential LF, 
because the number of galaxies in each group is small \citep{tre09}.
For the fainter galaxies ($M_V>-17$) in the NGC 2784 group,
we use the results from the surface photometry in Table \ref{tab-cat}.
For the three bright group galaxies, 
we also use the photometry obtained from our deep stack images:
NGC 2784 ($M_V = -20.85$), NGC 2835 ($M_V = -20.44$), and DDO 062 ($M_V = -17.83$).
These values are consistent to within 0.4 mag with the published total absolute magnitudes \citep{kar13}.
Because of increased contamination at larger projected radii, we use only 
the dwarf candidates within $R<0.5$ Mpc. 

To obtain the comparison LF's, we use photometry from the literature (for the NGC 5128 group from \citet{tul15} and \citet{crn16}; the 
  M81 group from \citet{chi09}; for the M83 group from \citet{mul15}; for the M106 group from \citet{kim11}, and for the Local Group from \citet{mcc12} and \citet{mar13a,mar13b}).
Where necessary, we transform magnitudes 
        using $M_V =  M_B-0.7$, 
        $M_V = M_r+0.2$ \citep{mul15}, and
         $V = g - 0.5784*(g - r) - 0.0038$ \citep{lup05}.

In
Figure \ref{fig-clf} we compare the cumulative LFs of six galaxy groups, including the NGC 2784 group. 
We fit cumulative Schechter functions \citep{sch76} to the observed cumulative LFs 
 according to the method described in Section 3.2 of \citet{chi09}.
The faint-end slopes ($\alpha$) of the fitted cumulative LFs
 have values ranging from $\alpha\approx-1.0$ to $\alpha\approx-1.4$ (Table \ref{tab-group}).
The slope ($\alpha=-1.33 \pm 0.05$) for the NGC 2784 group 
   is similar to those of the M81, M83, M106, and NGC 5218 groups, 
  while the slope is steeper than that of the Local Group 
  ($\alpha\approx -1.05$, which is consistent with the values from other studies \citep{chi09, mcc09, kim11}).
  The slopes for the NGC 2784 group and the other five groups are much flatter than the slope ($\alpha=-1.8$) of the  subhalo mass function in the $\Lambda$CDM model \citep{tre02}, shown as the dotted line in Figure \ref{fig-clf}. 
 We also check any dependency between the faint-end slope ($\alpha$) and the characteristic absolute magnitude ($M^*$) of the LF in each group.
The $\alpha$ values of six groups including the NGC 2784 group change slightly ($\lesssim 30$\%) within their uncertainties (e.g. as shown in Figure 30 by \citet{chi09}, 
Figure 16 by \citet{tre09}, and Figure 13 by \citet{fer16}).
Thus the $M^*$ is unlikely to affect significantly the $\alpha$ values estimated in this study.

We fit only over the magnitude range  $M_V<-10$ mag. We set the faint limit at 
a magnitude where we are still highly incomplete.   
The number of member galaxies used for the fits vary 
from 16 (for the M 106 group) to 55 (for the Local Group).
The variation in number is in large part due to the
more or less complete surveys of the relevant volumes, for example the 
survey area for M106 group extends only over $R<0.13$ Mpc \citep{kim11}, 
and to the different detection limits for dwarf galaxies with very low surface brightness or compact size.
Although the shape of the LF is not a function of the number of galaxies,  many of 
the potential reasons for incompleteness in the surveys could translate differentially at 
different magnitudes. For example, in NGC 2784 we found that brighter dwarfs are more
centrally concentrated, meaning that a survey that does not include the entire volume will
be biased toward brighter dwarfs (see Figure \ref{fig-rndenmag}). Likewise, incompleteness due to surface brightness effects will disfavor fainter galaxies.
As such, it is extremely difficult to compare results from
different surveys, which will have different explicit and implicit selection criteria. A major 
strength of our eventually larger KMTNet survey of dwarf galaxies will be the homogeneity of
the data and selection techniques. However, at the moment we can only compare results 
from various different surveys.

To search for a possible dependency of dwarf galaxy population with group properties, 
we investigate whether the faint-end slopes ($\alpha$) of LFs of the six groups vary 
  as a function of the group mass.
 We adopt two mass indicators (see Table \ref{tab-group}):
  the total $K$-band luminosity of the group galaxies ($L_K$),
  which correlates with the group mass \citep{mak11},
  and
  the velocity dispersion derived from the radial velocities 
   of the member galaxies in each group ($\sigma_{group}$).
  The galaxy luminosities and velocities are drawn from \citet{kar13}.
  We consider all galaxies associated with the same main disturber (MD) and $\Theta_1>0$ (tidal index) as members of each group (see \citet{kar13}).
In Figure \ref{fig-alpha} we show the relationships between $\alpha$ and total luminosity or velocity dispersion. There is a mild suggestion in both plots that 
$\alpha$ decreases (shallower slope) as the mass of the galaxy group ($L_K$ and $\sigma_{group}$) increases.
 When we simply fit the data with a linear function,
 we obtain $\alpha=0.38 {\rm log}(L_K)-5.47$ and $\alpha=0.002 \sigma_{group}-1.447$.
However, the correlation coefficient values (e.g.  Pearson's $r$ correlation, Spearman's rank correlation) estimated from the fits have approximately 1 sigma significance, and
 with this sample size statistical tests do not confirm this impression. 
In addition to the difficulty posed by the small statistical sample, we are concerned about 
systematic effects resulting 
from the comparison of results from diverse surveys. 
While intriguing, this test needs to be revisited once a large and uniform survey
is competed.
Our KMTNet survey of $15-20$ groups should be well positioned to address this 
topic.

\section{Summary}\label{summary}

Using  deep, wide field images obtained from the KMTNet Supernova Program, we identified 38 dwarf galaxy candidates, 31 previously unknown, in the nearby NGC 2784 galaxy group and measured their photometric properties.  KMTNet survey data produce competitive dwarf galaxy surveys to the current
state-of-the-art.
Our primary results are summarized as follows.

\begin{enumerate}

\item 
We discovered 31 dwarf galaxy candidates around NGC 2784 galaxy over 30 $\square^\circ$. 
Based on their cluster around NGC 2784, we estimate that at least 85\% of these are physically 
associated with the NGC 2784 group.
Beyond a projected separation of 0.5 Mpc the number density profile flattens and we may be detecting
only field contaminants. 

\item
We present surface photometry for 38 dwarfs, our new discoveries plus the previously known 7 dwarf galaxies \citep{kar13}.
Our $3\sigma$ surface brightnesses limit is approximately $\mu_V\sim 28.5$ mag arcsec$^{-2}$.
The central surface brightness and total absolute magnitude of the faintest dwarf galaxies we detect 
are $\mu_{0,V}\sim 26.1$  mag arcsec$^{-2}$ and $M_V\sim -9.5$ mag, respectively.
The effective radii of the dwarfs are all larger than $\sim 3.5$ arcsec ($
\sim 170$ pc at the distance of NGC 2784).

\item
The radial number density of the dwarf candidates decreases exponentially 
(or with power index, $\alpha \approx -1.3$) with distance from the center of NGC 2784.
The background contamination is about 15\%  and so estimate that about 27 of the 32 candidates within 2.7 degree are dwarf members of the group.

\item  
The color and S{\'e}rsic structural parameters for the dwarf candidates
  are consistent with those of dwarfs in other galaxy groups (M81, M83, M106,  and the Local Group).
The dwarf galaxy candidates have a color-magnitude relation similar to that of early-type galaxies.

\item  
The  color and number density radial profiles for the candidate dwarf population in the 
NGC 2784 group galaxies, shows that the central dwarfs are both brighter and redder than
the rest of the population. We caution that contamination is likely to produce 
relations of the same sense, but argue that the results are not dominated by the contamination.

\item 
The faint-end slope of the luminosity function (LF) of the NGC 2784 group galaxies 
  is $\alpha \approx -1.33\pm0.05$, 
  which is consistent with that seen in some other groups, but apparently 
  steeper than that of the Local Group galaxies. Eventually, the KMTNet survey will provide
  homogeneous catalogs of dwarf galaxies in 15 to 20 groups.

\end{enumerate}


\acknowledgments 
The authors are grateful to the staffs of the KMTNet. 
Drs. Youngdae Lee and Suk Kim are thanked for useful comments.
%


%

\clearpage

\begin{deluxetable}{ccccccc}
\tabletypesize{\tiny}
\tablewidth{0pc}
\tablecaption{Observing Log for deep stack images\label{tab-obslog1}}
\tablehead{
\colhead{Field} &
\colhead{Filter} &
\colhead{R.A.(J2000)} &
\colhead{Decl.(J2000)} &
\colhead{T(exp)} &
\colhead{Date(UT)} &
\colhead{Seeing} \\
\colhead{} &
\colhead{} &
\colhead{(hh:mm:ss)} &
\colhead{(dd:mm)} &
\colhead{(N$\times$sec)} &
\colhead{(2015.mm.dd.)} &
\colhead{(arcsec)} 
}
\startdata
N2784-3-Q0 & $I$ & 09:10:43 & -22:08 & 397 $\times$  60 s & 03.01. -- 06.30. &  1.4 \\ 
N2784-3-Q0 & $V$ & 09:10:43 & -22:08 & 383 $\times$  60 s & 03.01. -- 06.30. &  1.5 \\ 
N2784-3-Q0 & $B$ & 09:10:43 & -22:08 & 349 $\times$  60 s & 03.01. -- 06.30. &  1.6 \\ 
N2784-3-Q1 & $I$ & 09:06:05 & -22:08 & 397 $\times$  60 s & 03.01. -- 06.27. &  1.3 \\ 
N2784-3-Q1 & $V$ & 09:06:05 & -22:08 & 372 $\times$  60 s & 03.01. -- 06.30. &  1.4 \\ 
N2784-3-Q1 & $B$ & 09:06:05 & -22:08 & 345 $\times$  60 s & 03.01. -- 06.29. &  1.5 \\ 
N2784-3-Q2 & $I$ & 09:10:44 & -23:15 & 405 $\times$  60 s & 03.01. -- 06.30. &  1.3 \\ 
N2784-3-Q2 & $V$ & 09:10:44 & -23:15 & 395 $\times$  60 s & 03.01. -- 07.01. &  1.4 \\ 
N2784-3-Q2 & $B$ & 09:10:45 & -23:15 & 375 $\times$  60 s & 03.01. -- 06.30. &  1.5 \\ 
N2784-3-Q3 & $I$ & 09:06:03 & -23:15 & 390 $\times$  60 s & 03.01. -- 06.30. &  1.3 \\ 
N2784-3-Q3 & $V$ & 09:06:03 & -23:15 & 368 $\times$  60 s & 03.01. -- 06.30. &  1.4 \\ 
N2784-3-Q3 & $B$ & 09:06:03 & -23:15 & 346 $\times$  60 s & 03.01. -- 06.30. &  1.5 \\ 
N2784-4-Q0 & $I$ & 09:12:09 & -23:53 & 400 $\times$  60 s & 03.01. -- 07.03. &  1.4 \\ 
N2784-4-Q0 & $V$ & 09:12:09 & -23:53 & 375 $\times$  60 s & 03.01. -- 07.03. &  1.5 \\ 
N2784-4-Q0 & $B$ & 09:12:09 & -23:53 & 353 $\times$  60 s & 03.01. -- 07.01. &  1.7 \\ 
N2784-4-Q1 & $I$ & 09:07:27 & -23:53 & 400 $\times$  60 s & 03.01. -- 07.03. &  1.3 \\ 
N2784-4-Q1 & $V$ & 09:07:27 & -23:53 & 365 $\times$  60 s & 03.01. -- 07.03. &  1.4 \\ 
N2784-4-Q1 & $B$ & 09:07:27 & -23:53 & 350 $\times$  60 s & 03.01. -- 06.29. &  1.5 \\ 
N2784-4-Q2 & $I$ & 09:12:10 & -25:00 & 409 $\times$  60 s & 03.01. -- 07.03. &  1.3 \\ 
N2784-4-Q2 & $V$ & 09:12:10 & -25:00 & 385 $\times$  60 s & 03.01. -- 07.03. &  1.4 \\ 
N2784-4-Q2 & $B$ & 09:12:10 & -25:00 & 376 $\times$  60 s & 03.01. -- 07.03. &  1.5 \\ 
N2784-4-Q3 & $I$ & 09:07:25 & -25:00 & 394 $\times$  60 s & 03.01. -- 07.03. &  1.3 \\ 
N2784-4-Q3 & $V$ & 09:07:25 & -25:00 & 366 $\times$  60 s & 03.01. -- 07.03. &  1.4 \\ 
N2784-4-Q3 & $B$ & 09:07:26 & -25:00 & 350 $\times$  60 s & 03.01. -- 07.03. &  1.5 \\ 
N2784-5-Q0 & $I$ & 09:18:43 & -22:08 & 391 $\times$  60 s & 03.01. -- 07.01. &  1.4 \\ 
N2784-5-Q0 & $V$ & 09:18:43 & -22:08 & 361 $\times$  60 s & 03.01. -- 07.01. &  1.5 \\ 
N2784-5-Q0 & $B$ & 09:18:43 & -22:08 & 341 $\times$  60 s & 03.01. -- 07.01. &  1.7 \\ 
N2784-5-Q1 & $I$ & 09:14:04 & -22:08 & 386 $\times$  60 s & 03.01. -- 07.01. &  1.3 \\ 
N2784-5-Q1 & $V$ & 09:14:05 & -22:08 & 353 $\times$  60 s & 03.01. -- 07.01. &  1.4 \\ 
N2784-5-Q1 & $B$ & 09:14:05 & -22:08 & 331 $\times$  60 s & 03.01. -- 06.29. &  1.5 \\ 
N2784-5-Q2 & $I$ & 09:18:44 & -23:15 & 396 $\times$  60 s & 03.01. -- 07.01. &  1.3 \\ 
N2784-5-Q2 & $V$ & 09:18:45 & -23:15 & 375 $\times$  60 s & 03.01. -- 07.01. &  1.4 \\ 
N2784-5-Q2 & $B$ & 09:18:45 & -23:15 & 361 $\times$  60 s & 03.01. -- 07.01. &  1.5 \\ 
N2784-5-Q3 & $I$ & 09:14:03 & -23:15 & 383 $\times$  60 s & 03.01. -- 07.01. &  1.3 \\ 
N2784-5-Q3 & $V$ & 09:14:03 & -23:15 & 353 $\times$  60 s & 03.01. -- 07.01. &  1.4 \\ 
N2784-5-Q3 & $B$ & 09:14:03 & -23:15 & 337 $\times$  60 s & 03.01. -- 06.30. &  1.5 \\ 
N2784-6-Q0 & $I$ & 09:20:09 & -23:53 & 389 $\times$  60 s & 03.01. -- 06.30. &  1.4 \\ 
N2784-6-Q0 & $V$ & 09:20:09 & -23:53 & 361 $\times$  60 s & 03.01. -- 06.30. &  1.5 \\ 
N2784-6-Q0 & $B$ & 09:20:09 & -23:53 & 331 $\times$  60 s & 03.01. -- 06.30. &  1.6 \\ 
N2784-6-Q1 & $I$ & 09:15:27 & -23:53 & 383 $\times$  60 s & 03.01. -- 06.30. &  1.3 \\ 
N2784-6-Q1 & $V$ & 09:15:27 & -23:53 & 349 $\times$  60 s & 03.01. -- 06.30. &  1.4 \\ 
N2784-6-Q1 & $B$ & 09:15:27 & -23:53 & 328 $\times$  60 s & 03.01. -- 06.29. &  1.5 \\ 
N2784-6-Q2 & $I$ & 09:20:10 & -25:00 & 391 $\times$  60 s & 03.01. -- 06.30. &  1.3 \\ 
N2784-6-Q2 & $V$ & 09:20:10 & -25:00 & 366 $\times$  60 s & 03.01. -- 06.30. &  1.4 \\ 
N2784-6-Q2 & $B$ & 09:20:10 & -25:00 & 352 $\times$  60 s & 03.01. -- 06.30. &  1.5 \\ 
N2784-6-Q3 & $I$ & 09:15:26 & -25:00 & 377 $\times$  60 s & 03.01. -- 06.30. &  1.3 \\ 
N2784-6-Q3 & $V$ & 09:15:25 & -25:00 & 340 $\times$  60 s & 03.01. -- 06.30. &  1.4 \\ 
N2784-6-Q3 & $B$ & 09:15:26 & -25:00 & 324 $\times$  60 s & 03.01. -- 06.30. &  1.5 \\ 
N2784-7-Q0 & $I$ & 09:26:43 & -22:08 & 371 $\times$  60 s & 03.01. -- 06.29. &  1.4 \\ 
N2784-7-Q0 & $V$ & 09:26:43 & -22:08 & 346 $\times$  60 s & 03.01. -- 06.29. &  1.5 \\ 
N2784-7-Q0 & $B$ & 09:26:43 & -22:08 & 324 $\times$  60 s & 03.01. -- 06.29. &  1.7 \\ 
N2784-7-Q1 & $I$ & 09:22:05 & -22:08 & 371 $\times$  60 s & 03.01. -- 06.29. &  1.3 \\ 
N2784-7-Q1 & $V$ & 09:22:05 & -22:08 & 334 $\times$  60 s & 03.01. -- 06.29. &  1.4 \\ 
N2784-7-Q1 & $B$ & 09:22:05 & -22:08 & 317 $\times$  60 s & 03.01. -- 06.26. &  1.6 \\ 
N2784-7-Q2 & $I$ & 09:26:45 & -23:15 & 372 $\times$  60 s & 03.01. -- 06.29. &  1.3 \\ 
N2784-7-Q2 & $V$ & 09:26:45 & -23:15 & 352 $\times$  60 s & 03.01. -- 06.29. &  1.4 \\ 
N2784-7-Q2 & $B$ & 09:26:45 & -23:15 & 344 $\times$  60 s & 03.01. -- 06.29. &  1.5 \\ 
N2784-7-Q3 & $I$ & 09:22:03 & -23:15 & 363 $\times$  60 s & 03.01. -- 06.29. &  1.3 \\ 
N2784-7-Q3 & $V$ & 09:22:03 & -23:15 & 330 $\times$  60 s & 03.01. -- 06.29. &  1.4 \\ 
N2784-7-Q3 & $B$ & 09:22:03 & -23:15 & 316 $\times$  60 s & 03.01. -- 06.29. &  1.5 \\ 
\enddata
\end{deluxetable}
\clearpage

\begin{deluxetable}{ccccccc}
\tabletypesize{\tiny}
\tablewidth{0pc}
\tablecaption{Observing Log for shallow stack images\label{tab-obslog2}}
\tablehead{
\colhead{Field} &
\colhead{Filter} &
\colhead{R.A.(J2000)} &
\colhead{Decl.(J2000)} &
\colhead{T(exp)} &
\colhead{Date(UT)} &
\colhead{Seeing} \\
\colhead{} &
\colhead{} &
\colhead{(hh:mm:ss)} &
\colhead{(dd:mm)} &
\colhead{(N$\times$sec)} &
\colhead{(2015.mm.dd.)} &
\colhead{(arcsec)} 
}
\startdata
N2784-1-Q0 & $I$ & 09:03:22 & -21:56 &   8 $\times$ 100 s & 02.09. -- 02.27. &  1.3 \\ 
N2784-1-Q0 & $V$ & 09:03:22 & -21:56 &   7 $\times$ 100 s & 02.09. -- 02.27. &  1.2 \\ 
N2784-1-Q0 & $B$ & 09:03:22 & -21:56 &   6 $\times$ 100 s & 02.09. -- 02.27. &  1.4 \\ 
N2784-1-Q1 & $I$ & 08:58:44 & -21:56 &   8 $\times$ 100 s & 02.09. -- 02.27. &  1.3 \\ 
N2784-1-Q1 & $V$ & 08:58:44 & -21:56 &   7 $\times$ 100 s & 02.09. -- 02.27. &  1.2 \\ 
N2784-1-Q1 & $B$ & 08:58:44 & -21:56 &   8 $\times$ 100 s & 02.09. -- 02.27. &  1.5 \\ 
N2784-1-Q2 & $I$ & 09:03:23 & -23:03 &   8 $\times$ 100 s & 02.09. -- 02.27. &  1.2 \\ 
N2784-1-Q2 & $V$ & 09:03:23 & -23:03 &   7 $\times$ 100 s & 02.09. -- 02.27. &  1.2 \\ 
N2784-1-Q2 & $B$ & 09:03:23 & -23:03 &   6 $\times$ 100 s & 02.09. -- 02.27. &  1.2 \\ 
N2784-1-Q3 & $I$ & 08:58:42 & -23:03 &   8 $\times$ 100 s & 02.09. -- 02.27. &  1.2 \\ 
N2784-1-Q3 & $V$ & 08:58:43 & -23:03 &   7 $\times$ 100 s & 02.09. -- 02.27. &  1.2 \\ 
N2784-1-Q3 & $B$ & 08:58:43 & -23:03 &   6 $\times$ 100 s & 02.09. -- 02.27. &  1.3 \\ 
N2784-2-Q0 & $I$ & 09:03:24 & -23:56 &   5 $\times$ 100 s & 02.09. -- 02.27. &  1.0 \\ 
N2784-2-Q0 & $V$ & 09:03:24 & -23:56 &   5 $\times$ 100 s & 02.09. -- 02.27. &  1.2 \\ 
N2784-2-Q0 & $B$ & 09:03:24 & -23:56 &   5 $\times$ 100 s & 02.09. -- 02.27. &  1.2 \\ 
N2784-2-Q1 & $I$ & 08:58:42 & -23:56 &   5 $\times$ 100 s & 02.09. -- 02.27. &  1.1 \\ 
N2784-2-Q1 & $V$ & 08:58:42 & -23:56 &   5 $\times$ 100 s & 02.09. -- 02.27. &  1.2 \\ 
N2784-2-Q1 & $B$ & 08:58:42 & -23:56 &   4 $\times$ 100 s & 02.09. -- 02.27. &  1.3 \\ 
N2784-2-Q2 & $I$ & 09:03:25 & -25:03 &   5 $\times$ 100 s & 02.09. -- 02.27. &  1.0 \\ 
N2784-2-Q2 & $V$ & 09:03:25 & -25:03 &   5 $\times$ 100 s & 02.09. -- 02.27. &  1.2 \\ 
N2784-2-Q2 & $B$ & 09:03:26 & -25:03 &   5 $\times$ 100 s & 02.09. -- 02.27. &  1.2 \\ 
N2784-2-Q3 & $I$ & 08:58:40 & -25:03 &   5 $\times$ 100 s & 02.09. -- 02.27. &  1.1 \\ 
N2784-2-Q3 & $V$ & 08:58:40 & -25:03 &   5 $\times$ 100 s & 02.09. -- 02.27. &  1.2 \\ 
N2784-2-Q3 & $B$ & 08:58:40 & -25:03 &   5 $\times$ 100 s & 02.09. -- 02.27. &  1.3 \\ 
N2784-8-Q0 & $I$ & 09:28:09 & -23:53 &   4 $\times$ 100 s & 02.11. -- 02.15. &  1.1 \\ 
N2784-8-Q0 & $V$ & 09:28:09 & -23:53 &   3 $\times$ 100 s & 02.11. -- 02.15. &  1.1 \\ 
N2784-8-Q0 & $B$ & 09:28:09 & -23:53 &   4 $\times$ 100 s & 02.11. -- 02.15. &  1.3 \\ 
N2784-8-Q1 & $I$ & 09:23:27 & -23:53 &   4 $\times$ 100 s & 02.11. -- 02.15. &  1.1 \\ 
N2784-8-Q1 & $V$ & 09:23:27 & -23:53 &   3 $\times$ 100 s & 02.11. -- 02.15. &  1.1 \\ 
N2784-8-Q1 & $B$ & 09:23:27 & -23:53 &   4 $\times$ 100 s & 02.11. -- 02.15. &  1.3 \\ 
N2784-8-Q2 & $I$ & 09:28:10 & -25:00 &   4 $\times$ 100 s & 02.11. -- 02.15. &  1.3 \\ 
N2784-8-Q2 & $V$ & 09:28:10 & -25:00 &   3 $\times$ 100 s & 02.11. -- 02.15. &  1.1 \\ 
N2784-8-Q2 & $B$ & 09:28:10 & -25:00 &   4 $\times$ 100 s & 02.11. -- 02.15. &  1.3 \\ 
N2784-8-Q3 & $I$ & 09:23:25 & -25:00 &   4 $\times$ 100 s & 02.11. -- 02.15. &  1.2 \\ 
N2784-8-Q3 & $V$ & 09:23:25 & -25:00 &   3 $\times$ 100 s & 02.11. -- 02.15. &  1.1 \\ 
N2784-8-Q3 & $B$ & 09:23:26 & -25:00 &   4 $\times$ 100 s & 02.11. -- 02.15. &  1.3 \\ 
\enddata
\end{deluxetable}
\clearpage

\begin{deluxetable}{lccccccrrr}
\rotate
\tabletypesize{\tiny} 
\tablewidth{0pc}
\tablecaption{Photometric Catalog for Dwarf Galaxy Candidates in N2784 field\label{tab-cat}}
\tablehead{
\colhead{ID$~^a$} &
\colhead{R.A.(J2000)} &
\colhead{Dec.(J2000)} &
\colhead{$I~^b$} &
\colhead{$(V-I)~^c$} &
\colhead{$(B-V)~^c$} &
\colhead{$\mu_{0,I}~^d$} &
\colhead{$r_{0,I}~^d$} &
\colhead{$n_{I}~^d$} &
\colhead{$r_{e,I}~^d$} 
\\
\colhead{} &
\colhead{(hh:mm:ss)} &
\colhead{(dd:mm:ss)} &
\colhead{(mag)} &
\colhead{(mag)} &
\colhead{(mag)} &
\colhead{(mag arcsec$^{-2}$)} &
\colhead{(arcsec)} &
\colhead{} &
\colhead{(arcsec)} 
}
\startdata
         KSP-DW1 & 08:56:52.2 & -24:58:05 &  17.07 &   0.84 &   0.75 &  22.95 $\pm$   0.16 &   6.05 $\pm$   1.01 &  0.98 $\pm$   0.14  &   10.00    \\ 
         KSP-DW2 & 08:58:04.3 & -21:51:02 &  16.72 &   0.50 &   0.93 &  22.32 $\pm$   0.15 &   4.29 $\pm$   0.79 &  1.20 $\pm$   0.15  &    7.87    \\ 
         KSP-DW3 & 08:58:25.5 & -24:44:55 &  16.17 &   0.87 &   0.91 &  22.20 $\pm$   0.03 &   8.14 $\pm$   0.24 &  0.69 $\pm$   0.04  &    8.88    \\ 
         KSP-DW4 & 08:58:29.5 & -25:06:18 &  15.52 &   1.09 &   0.83 &  22.71 $\pm$   0.06 &  11.98 $\pm$   0.68 &  0.90 $\pm$   0.06  &   18.32    \\ 
         KSP-DW5 & 08:59:59.3 & -25:16:02 &  15.38 &   0.68 &   0.87 &  21.84 $\pm$   0.04 &   5.95 $\pm$   0.35 &  1.27 $\pm$   0.05  &   11.11    \\ 
      ESO497-004 & 09:03:03.1 & -23:48:24 &  13.75 &   0.56 &   0.74 &  21.36 $\pm$   0.01 &  15.44 $\pm$   0.23 &  0.82 $\pm$   0.01  &   21.47    \\ 
         KSP-DW6 & 09:04:02.4 & -22:56:52 &  18.66 &   1.13 &   ...  &  25.16 $\pm$   0.04 &  11.04 $\pm$   0.32 &  0.54 $\pm$   0.06  &    6.69    \\ 
         KSP-DW7 & 09:04:14.7 & -24:09:11 &  18.23 &   0.78 &   0.61 &  23.19 $\pm$   0.21 &   4.92 $\pm$   0.82 &  0.72 $\pm$   0.19  &    5.66    \\ 
         KSP-DW8 & 09:06:59.6 & -25:07:50 &  17.62 &   0.66 &   0.79 &  22.68 $\pm$   0.04 &   4.94 $\pm$   0.17 &  0.77 $\pm$   0.03  &    6.32   \\
         KSP-DW9 & 09:07:15.3 & -24:11:30 &  18.58 &   0.72 &   0.94 &  23.77 $\pm$   0.05 &   5.14 $\pm$   0.26 &  0.80 $\pm$   0.05  &    6.98   \\ 
        KSP-DW10 & 09:08:01.9 & -25:15:18 &  17.66 &   0.53 &   0.71 &  23.06 $\pm$   0.05 &   5.39 $\pm$   0.26 &  0.86 $\pm$   0.04  &    7.93   \\
        KSP-DW11 & 09:08:53.3 & -23:08:33 &  17.15 &   0.46 &   0.71 &  23.96 $\pm$   0.04 &  12.18 $\pm$   0.34 &  0.62 $\pm$   0.02  &   10.49    \\ 
        KSP-DW12 & 09:09:06.4 & -23:53:58 &  19.67 &   0.58 &   0.74 &  24.07 $\pm$   0.21 &   3.56 $\pm$   0.65 &  0.81 $\pm$   0.15  &    4.87    \\ 
            KK71 & 09:09:10.9 & -23:22:01 &  16.26 &   0.68 &   0.94 &  23.22 $\pm$   0.02 &  11.12 $\pm$   0.18 &  0.85 $\pm$   0.02  &   16.15    \\ 
          DDO056 & 09:09:45.9 & -23:00:30 &  13.76 &   0.68 &   0.87 &  22.07 $\pm$   0.01 &  23.43 $\pm$   0.12 &  0.69 $\pm$   0.00  &   24.96    \\ 
        KSP-DW13 & 09:10:00.0 & -24:10:19 &  16.85 &   0.52 &   1.09 &  22.86 $\pm$   0.07 &   4.30 $\pm$   0.39 &  1.37 $\pm$   0.07  &    8.17   \\ 
        KSP-DW14 & 09:10:25.0 & -24:01:30 &  16.69 &   0.78 &   0.97 &  23.56 $\pm$   0.07 &   7.03 $\pm$   0.58 &  1.29 $\pm$   0.06  &   13.19    \\ 
        KSP-DW15 & 09:11:10.7 & -24:24:41 &  16.56 &   0.77 &   1.06 &  23.86 $\pm$   0.09 &  11.78 $\pm$   1.02 &  0.97 $\pm$   0.08  &   19.28   \\ 
        KSP-DW16 & 09:11:26.9 & -23:08:44 &  18.20 &   0.56 &   0.77 &  23.57 $\pm$   0.02 &   6.48 $\pm$   0.12 &  0.57 $\pm$   0.02  &    4.41    \\ 
        KSP-DW17 & 09:11:36.2 & -23:30:00 &  15.52 &   0.61 &   0.88 &  22.14 $\pm$   0.02 &   8.86 $\pm$   0.19 &  0.94 $\pm$   0.02  &   14.14    \\ 
        KSP-DW18 & 09:11:42.2 & -24:13:29 &  18.59 &   0.42 &   0.80 &  23.75 $\pm$   0.10 &   4.79 $\pm$   0.48 &  0.87 $\pm$   0.11  &    7.10    \\ 
            KK72 & 09:11:42.4 & -23:58:56 &  16.88 &   0.73 &   0.96 &  23.93 $\pm$   0.02 &  14.17 $\pm$   0.20 &  0.55 $\pm$   0.02  &    8.89   \\ 
    NGC2784 dw01 & 09:12:17.0 & -24:12:47 &  15.38 &   0.83 &   0.99 &  23.54 $\pm$   0.02 &  21.17 $\pm$   0.31 &  0.74 $\pm$   0.02  &   25.58   \\ 
            KK73 & 09:12:29.2 & -24:14:28 &  14.45 &   0.82 &   1.08 &  21.90 $\pm$   0.03 &   9.46 $\pm$   0.30 &  1.26 $\pm$   0.02  &   17.62    \\ 
        KSP-DW19 & 09:14:32.4 & -23:33:26 &  15.26 &   0.53 &   0.73 &  22.68 $\pm$   0.04 &  12.81 $\pm$   0.39 &  0.94 $\pm$   0.02  &   20.45    \\ 
        KSP-DW20 & 09:14:40.5 & -24:15:10 &  17.74 &   0.71 &   0.89 &  23.82 $\pm$   0.03 &   7.50 $\pm$   0.26 &  0.84 $\pm$   0.03  &   10.80    \\ 
        KSP-DW21 & 09:15:08.4 & -25:02:58 &  20.15 &   0.79 &   0.88 &  23.87 $\pm$   0.40 &   2.26 $\pm$   0.81 &  0.97 $\pm$   0.25  &    3.71    \\ 
        KSP-DW22 & 09:15:35.3 & -23:48:57 &  17.57 &   0.33 &   0.64 &  23.47 $\pm$   0.03 &   7.54 $\pm$   0.23 &  0.72 $\pm$   0.03  &    8.81    \\ 
        KSP-DW23 & 09:16:01.6 & -23:36:23 &  20.47 &   ...  &   ...  &  24.33 $\pm$   0.59 &   2.02 $\pm$   1.24 &  1.15 $\pm$   0.46  &    3.64    \\ 
        KSP-DW24 & 09:16:08.5 & -23:43:17 &  18.11 &   0.72 &   0.83 &  24.04 $\pm$   0.04 &   7.47 $\pm$   0.30 &  0.75 $\pm$   0.05  &    9.27    \\ 
        KSP-DW25 & 09:16:41.5 & -23:06:08 &  20.19 &   0.47 &   1.21 &  24.75 $\pm$   0.08 &   4.68 $\pm$   0.24 &  0.46 $\pm$   0.06  &    3.63    \\ 
HIPASS J0916-23b & 09:16:58.3 & -23:16:45 &  14.08 &   0.33 &   0.57 &  20.78 $\pm$   0.01 &   9.25 $\pm$   0.18 &  0.93 $\pm$   0.01  &   14.65    \\ 
        KSP-DW26 & 09:17:17.3 & -24:22:05 &  19.83 &   0.42 &   1.03 &  24.80 $\pm$   0.08 &   5.37 $\pm$   0.32 &  0.57 $\pm$   0.07  &    3.81    \\ 
        KSP-DW27 & 09:17:24.8 & -22:33:42 &  18.48 &   1.18 &   1.16 &  25.04 $\pm$   0.06 &  11.27 $\pm$   0.44 &  0.56 $\pm$   0.09  &    7.30    \\ 
        KSP-DW28 & 09:17:43.4 & -23:29:37 &  16.64 &   0.49 &   0.74 &  23.18 $\pm$   0.02 &   9.71 $\pm$   0.19 &  0.78 $\pm$   0.02  &   12.72    \\ 
        KSP-DW29 & 09:20:08.2 & -25:00:06 &  19.19 &   0.67 &   0.87 &  23.61 $\pm$   0.06 &   3.85 $\pm$   0.20 &  0.70 $\pm$   0.05  &    4.25    \\ 
        KSP-DW30 & 09:20:13.2 & -23:56:02 &  18.11 &   0.61 &   0.54 &  24.37 $\pm$   0.04 &   8.12 $\pm$   0.33 &  0.85 $\pm$   0.07  &   11.72 \\ 
        KSP-DW31 & 09:23:44.0 & -23:03:57 &  19.39 &   0.71 &   0.62 &  24.57 $\pm$   0.07 &   5.62 $\pm$   0.34 &  0.66 $\pm$   0.09  &    5.60    \\ 
\enddata
\tablenotetext{~}{
$^a$ KSP-DW** condidates are newly discovered objects in KSP, while other names are from \citet{kar13}. 
$^b$ $I$ is $I$-band total magnitude derived from S{\'e}rsic fit.
$^c$ $(V-I)$ and $(B-V)$ are colors without the extinction correction. 
$^d$ $\mu_{0,I}$, $r_{0,I}$, $n_{I}$, and $r_{e,I}$ are central surface brightness, scale length, curvature index, and effective radius derived for the $I$-band  S{\'e}rsic fits, respectively.
}
\end{deluxetable}
\clearpage

\begin{deluxetable}{lccr}
\tablewidth{0pc}
\tablecaption{Faint-end slopes of the luminosity function of galaxy groups  \label{tab-group}}
\tablehead{
\colhead{Group} &
\colhead{$\alpha$} &
\colhead{log $L_{K}$} &
\colhead{$\sigma_{group}$}
\\
\colhead{} &
\colhead{} &
\colhead{($L_{\sun}$)} &
\colhead{(km s$^{-1}$)} 
}
\startdata
     M83      &  $-1.39 \pm$  0.05 & 10.87 &   60 $\pm$  18 \\
     NGC 2784 &  $-1.33 \pm$  0.05 & 10.91 &  107 $\pm$  61 \\
     M106     &  $-1.20 \pm$  0.10 & 10.97 &  127 $\pm$  36 \\
 Local Group  &  $-1.05 \pm$  0.08 & 11.07 &  108 $\pm$  13 \\
       M81    &  $-1.31 \pm$  0.06 & 11.11 &  119 $\pm$  23 \\
     NGC 5128 &  $-1.27 \pm$  0.05 & 11.17 &  137 $\pm$  34 \\
\enddata
\end{deluxetable}
\clearpage

\clearpage

\clearpage

\clearpage

\begin{figure}
\epsscale{1.0} 
\plotone{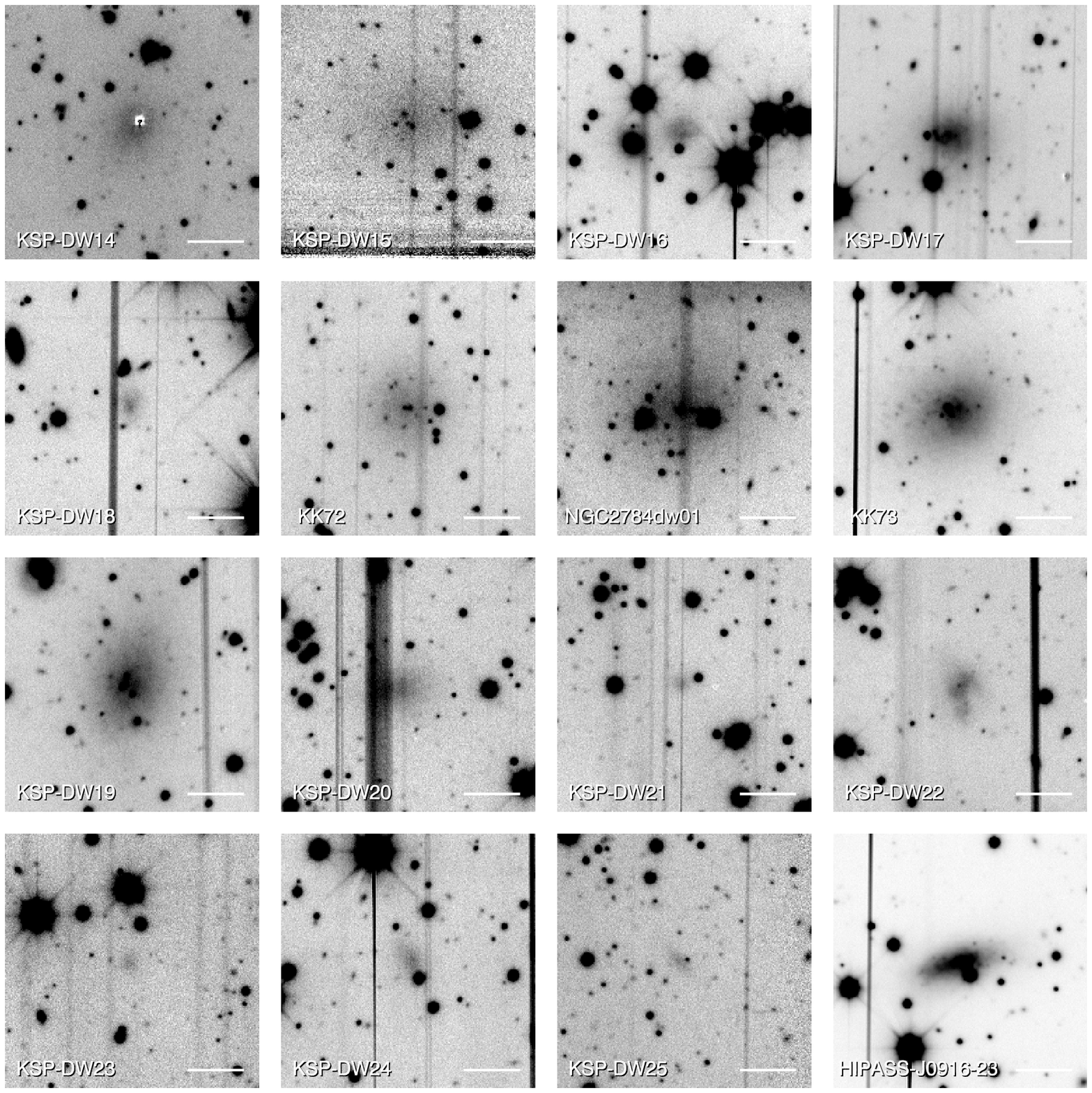}
\centerline{Fig. 2. --- {\it Continued}}
\end{figure}
\clearpage

\begin{figure}
\epsscale{1.0} 
\plotone{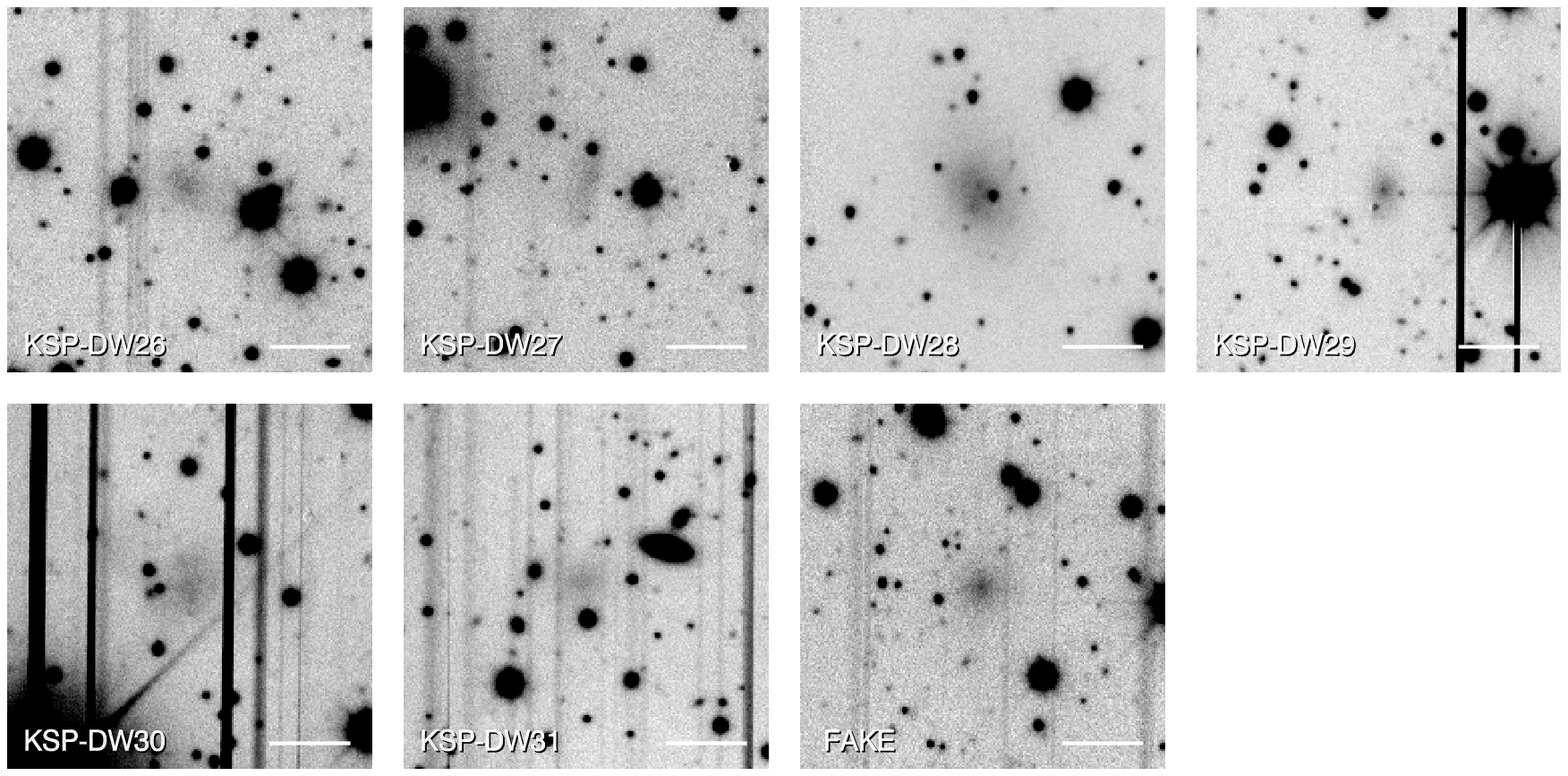}
\centerline{Fig. 2. --- {\it Continued}}
\end{figure}
\clearpage

\begin{figure}
\epsscale{1.0}
\plotone{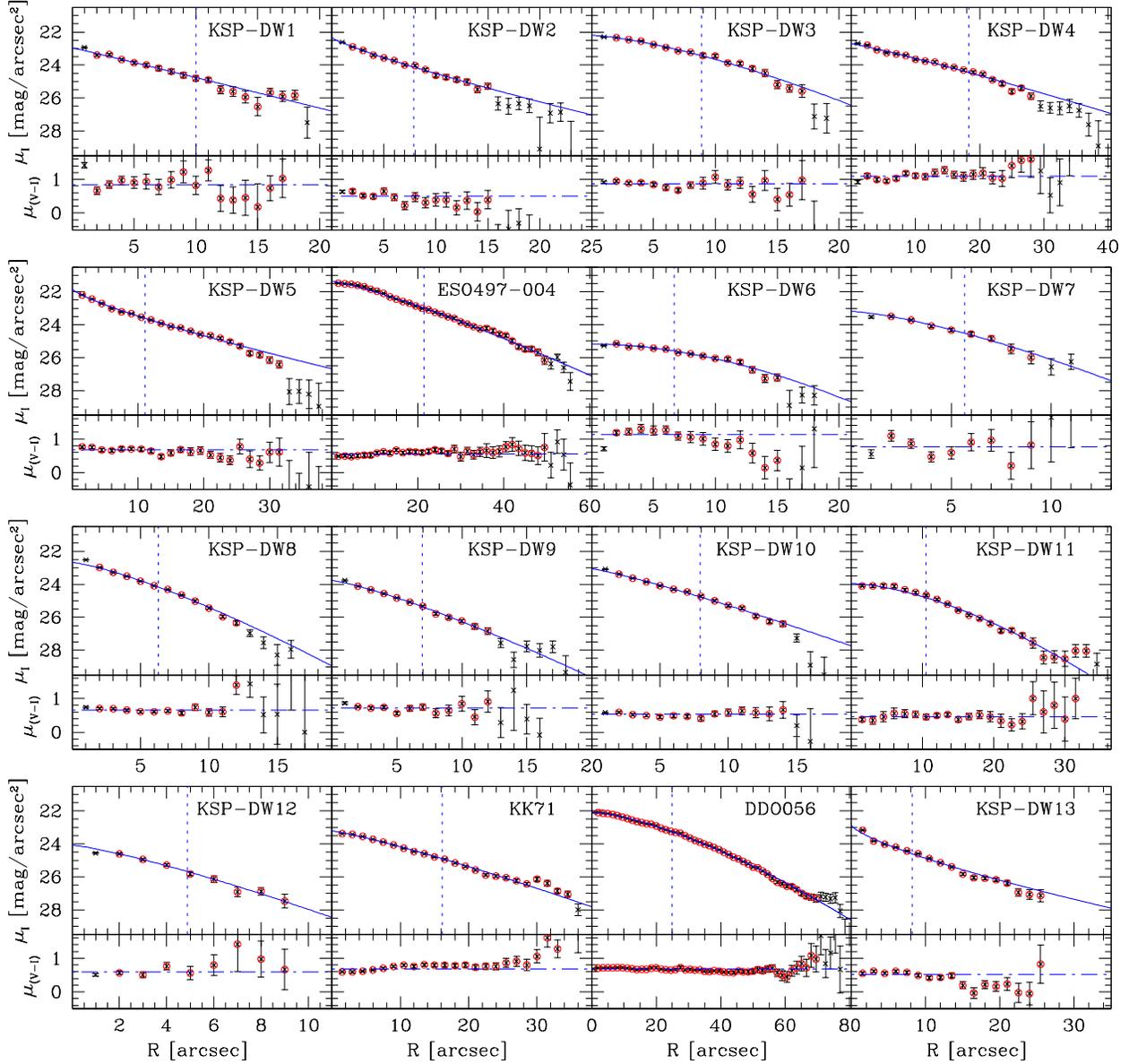}
\caption{Surface brightness and color profiles of the dwarf candidates. 
The open circles in each panel indicate surface brightnesses and colors above approximately 3 sigma sky level.
   They are used for S{\'e}rsic fit and mean color estimation represented by the solid and dot-dashed lines, respectively.
    The vertical dotted lines represent effective radius.
\label{fig-radimu0a}}
\end{figure}
\clearpage

\begin{figure}
\epsscale{1.0}
\plotone{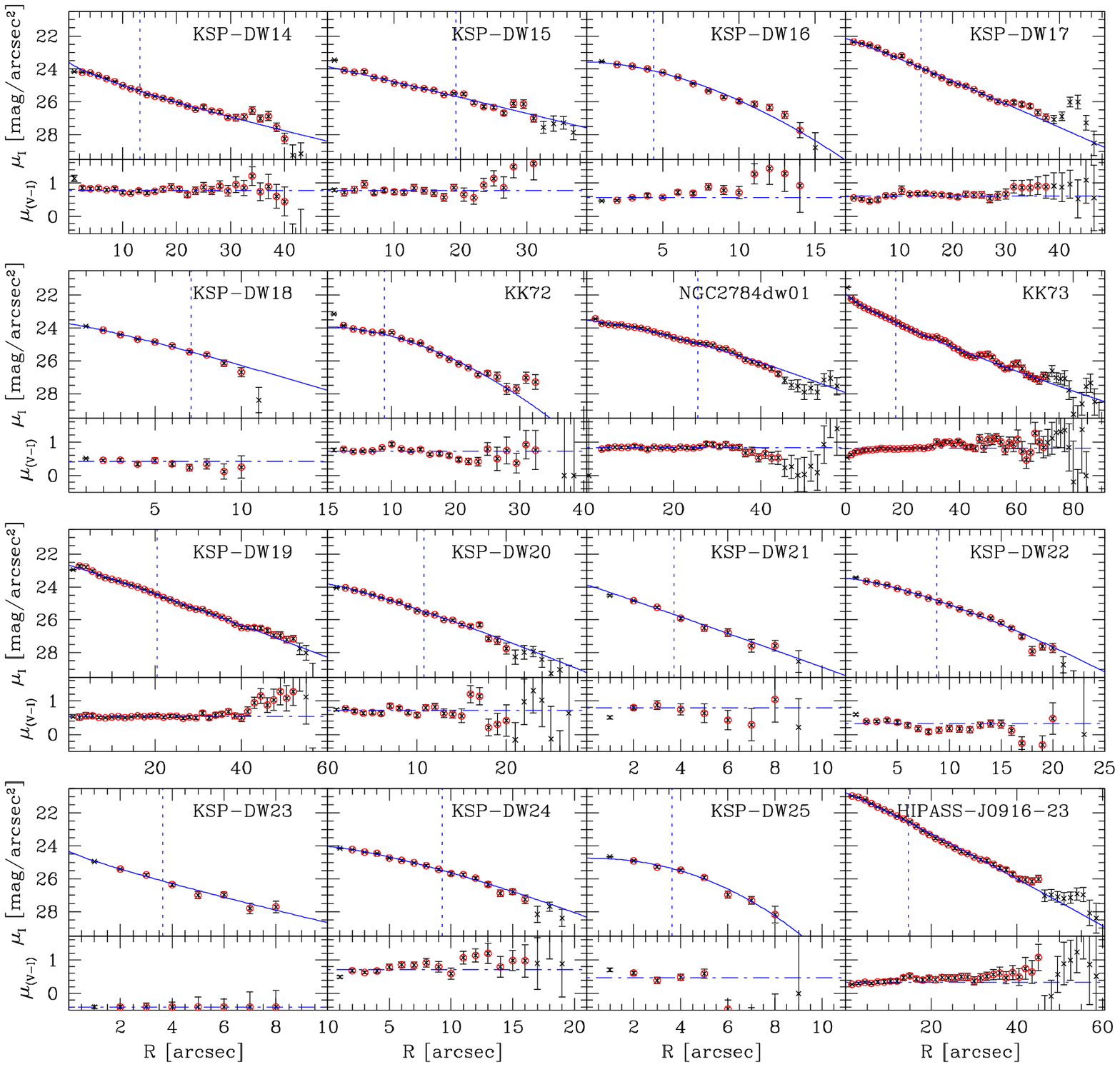}
\centerline{Fig. 3. --- {\it Continued}}
\end{figure}
\clearpage

\begin{figure}
\epsscale{1.0}
\plotone{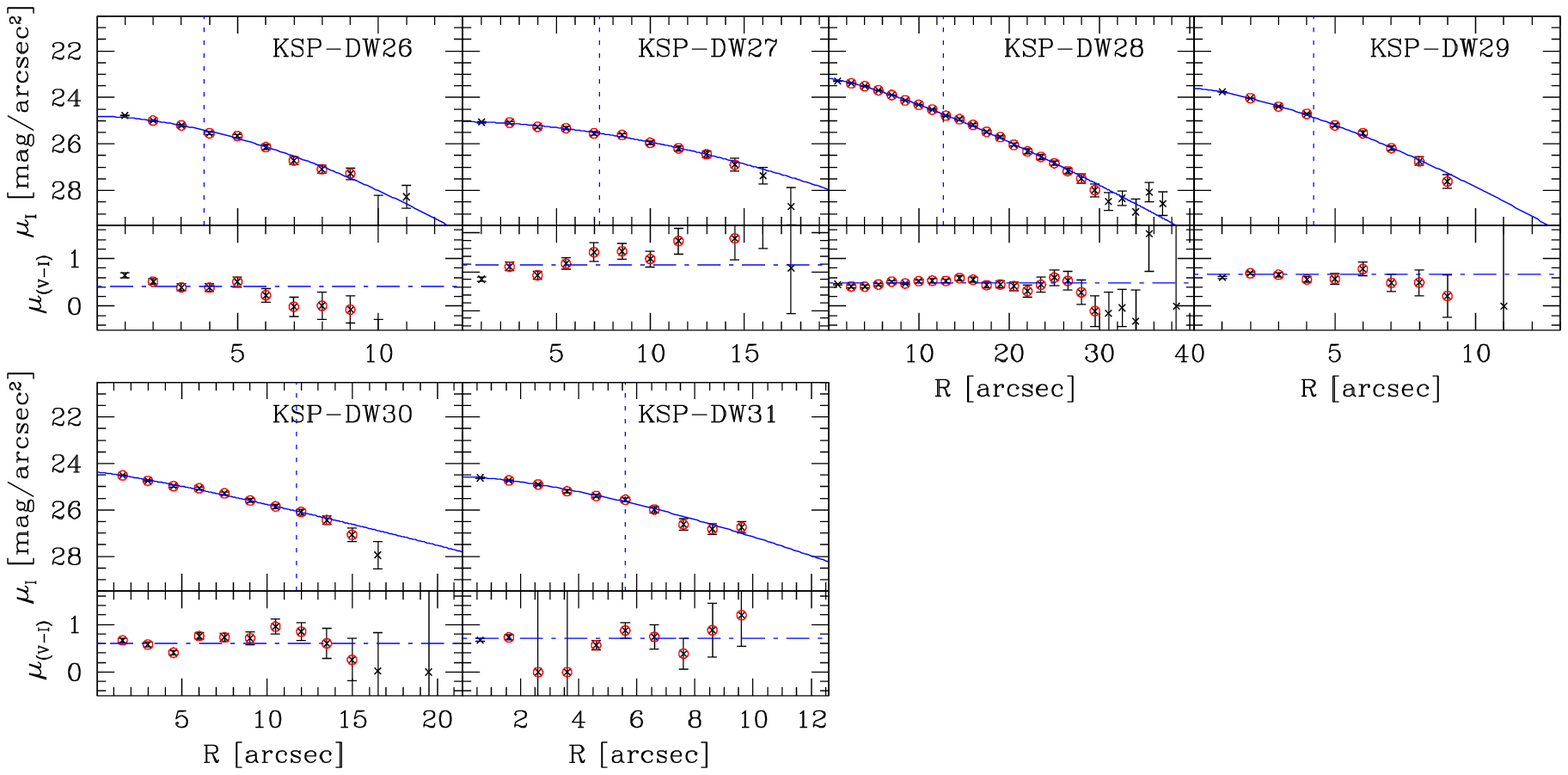}
\centerline{Fig. 3. --- {\it Continued}}
\end{figure}
\clearpage

\begin{figure}
\epsscale{0.9} 
\plotone{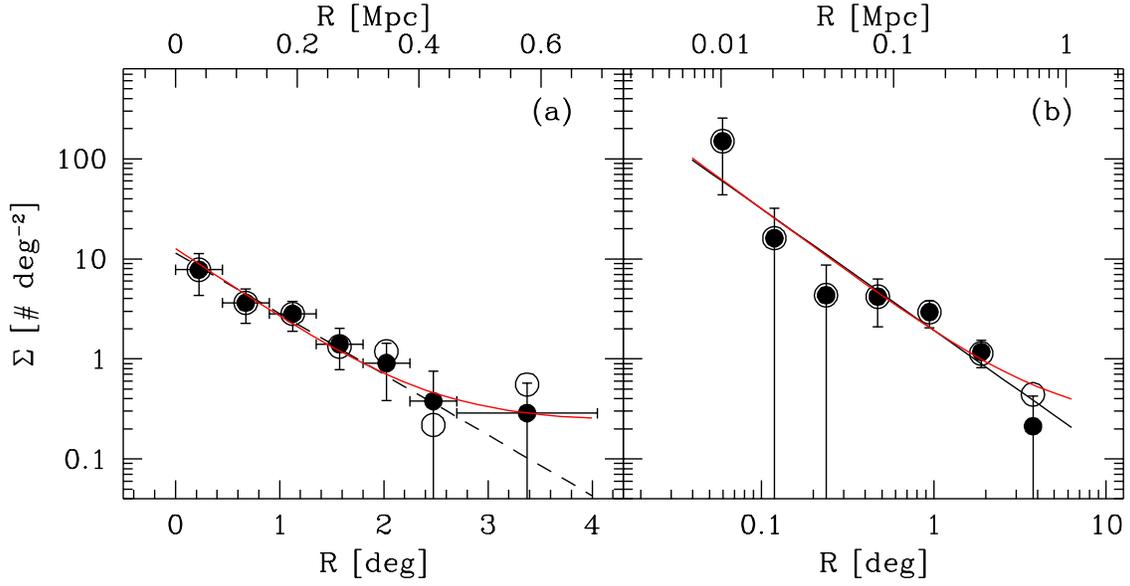}
\caption{Radial number density of the dwarf galaxy candidates in the NGC 2784 field.
The filled circles indicate the number densities obtained from the visually identified dwarf galaxies using only the region covered by deep stack images, while the open circles when using the region covered by both shallow and deep stacks.
(a) The radial bin has equal linear interval except the last bin. 
The dashed line and solid curve indicate the exponential fit with $R\lesssim3$ deg and
 the exponential function + $constant$ fit, respectively.
(b) The radial bin has equal logarithmic interval.
The solid line and curve indicate the power-law fits without and with $constant$ value, respectively.
\label{fig-rnden}}
\end{figure}
\clearpage

\begin{figure}
\epsscale{0.6}
\plotone{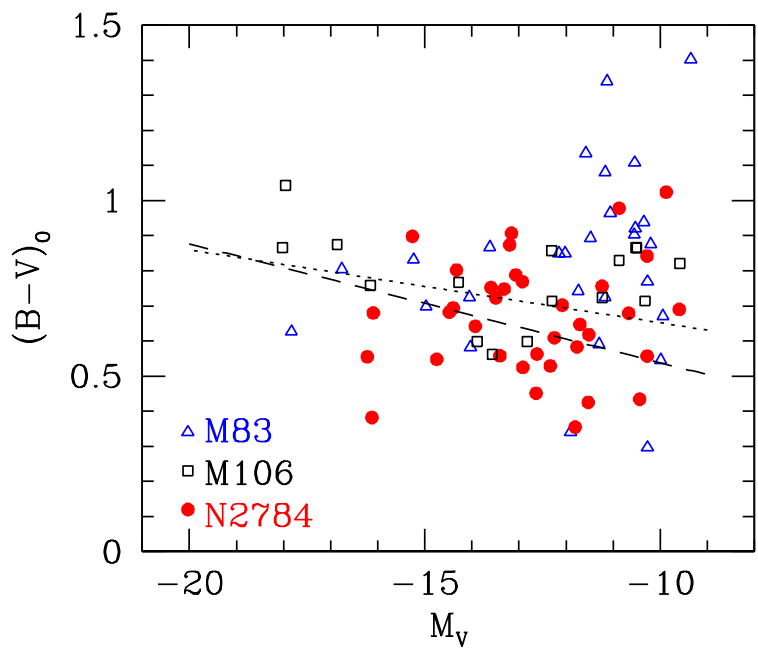}
\caption{ 
Color-Magnitude diagram of the dwarf galaxy candidates in NGC 2784.
The circles, triangles, and boxes represent the galaxy candidates 
in the NGC 2784 field (this study), M83 group \citep{mul15}, and M106 group \citep{kim11}, respectively. 
The dashed line denotes the color-magnitude relation obtained from the early-type galaxies in less massive and less dense Ursa Major cluster \citep{pak14}, while
the dotted line from the early-type dwarf galaxies in more massive and more dense Virgo cluster \citep{lis08}.
\label{fig-cmd}}
\end{figure}
\clearpage

\begin{figure}
\epsscale{0.55}
\plotone{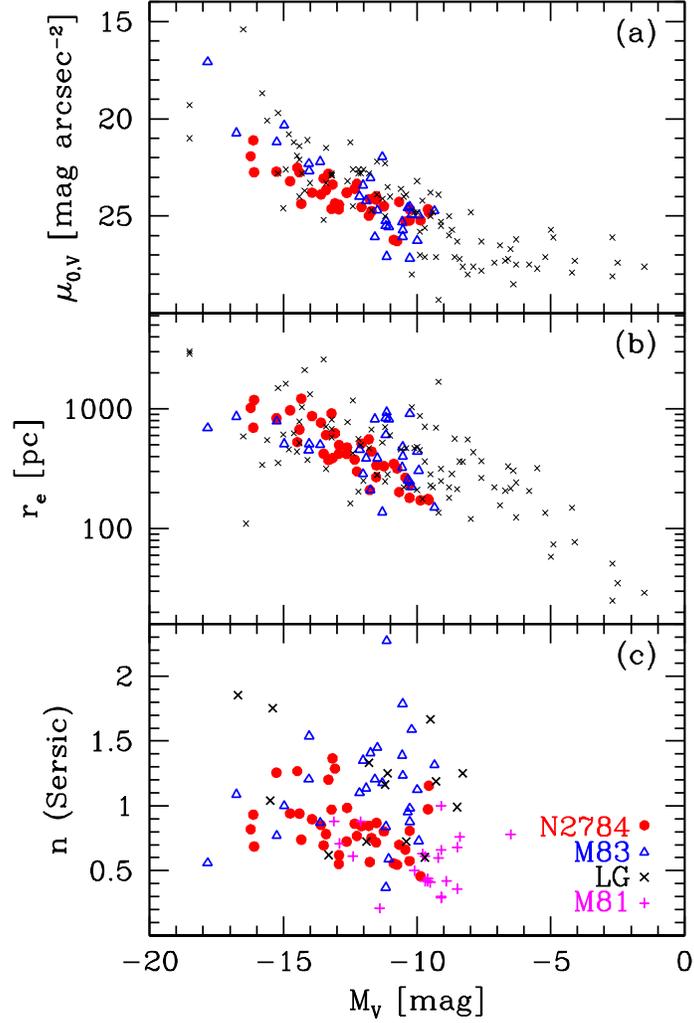}
\caption{Central surface brightness (a), effective radius (b), and S{\'e}rsic-$n$ (c) versus total absolute magnitudes of the dwarf galaxies.
The circles, triangles, crosses, and pluses indicate the dwarfs 
in the NGC 2784 field (this study), M83 group \citep{mul15},  
 the Local Group (\citet{mcc12} for $\mu_0, r_{e}$ and \citet{jer00} for S{\'e}rsic-$n$), 
 and  M81 group \citep{chi09},
respectively. 
\label{fig-censurface}}
\end{figure}
\clearpage

\begin{figure}
\epsscale{0.6}
\plotone{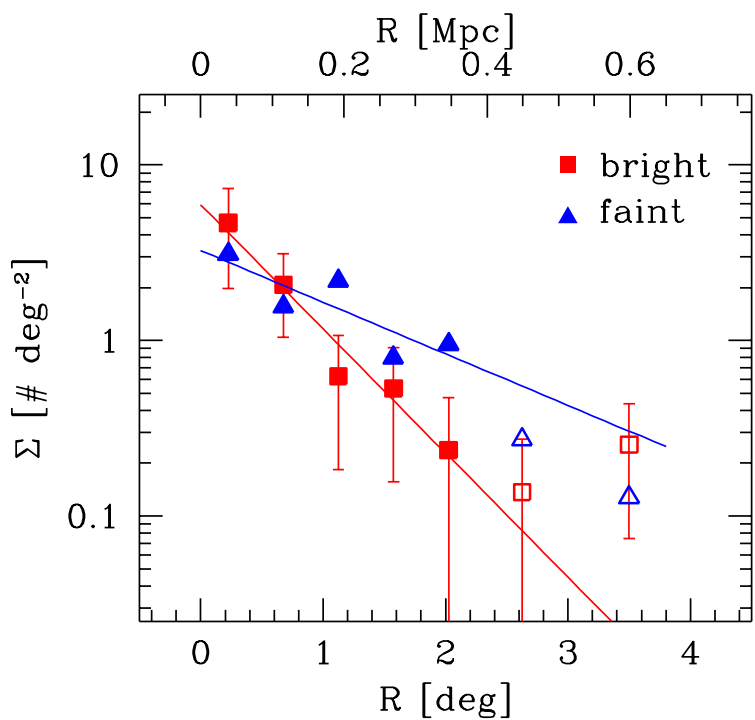}
\caption{Radial number density of the dwarf galaxy candidates in the NGC 2784 field.
The boxes and triangles are the bright ($M_V<-13$) and faint ($M_V\geq-13$) dwarfs, respectively.
The solid lines represent exponential fits to each of the group.
The filled and open symbols indicate inner and outer samples divided at $R\approx 2.5^\circ$, respectively.
The filled symbols, mostly derived from the deep stack images, are used to get the fits.
\label{fig-rndenmag}}
\end{figure}
\clearpage

\begin{figure}
\epsscale{0.5}
\plotone{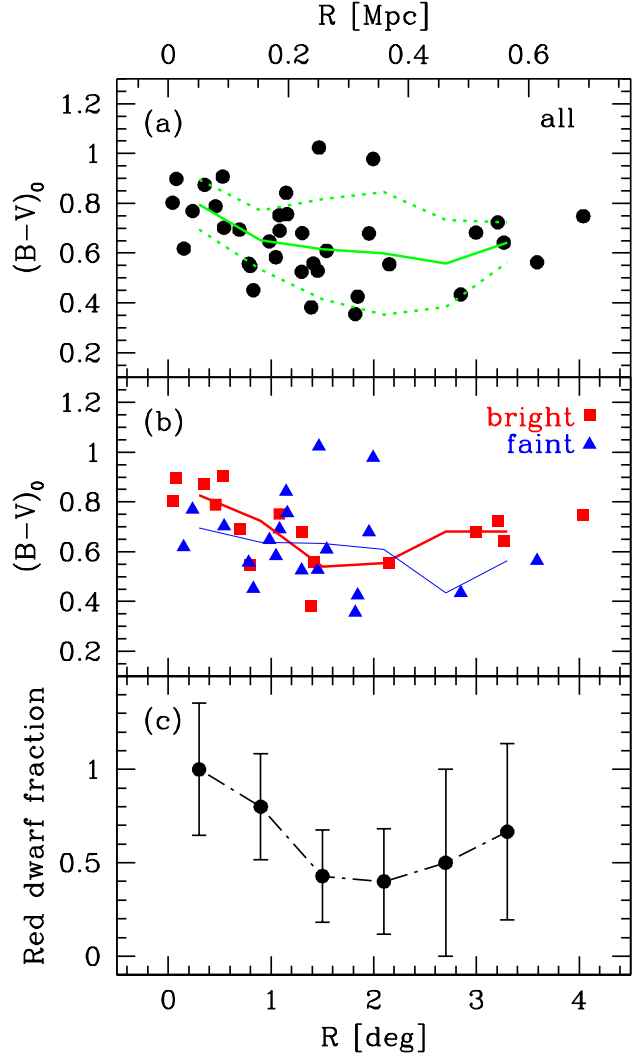}
\caption{Radial color distributions (a and b) and red dwarf fraction (c) of the dwarf galaxy candidates in the NGC 2784 field.
The solid and dotted curves indicate the mean color values and standard deviation in each radial bin, respectively.
The circles, boxes, and triangles represent all, bright ($M_V<-13$), and faint ($M_V\geq-13$) dwarf samples, respectively.
The errors of red dwarf fraction are estimated by assuming a Poisson distribution.
\label{fig-rcolor}}
\end{figure}
\clearpage

\begin{figure}
\epsscale{0.8}
\plotone{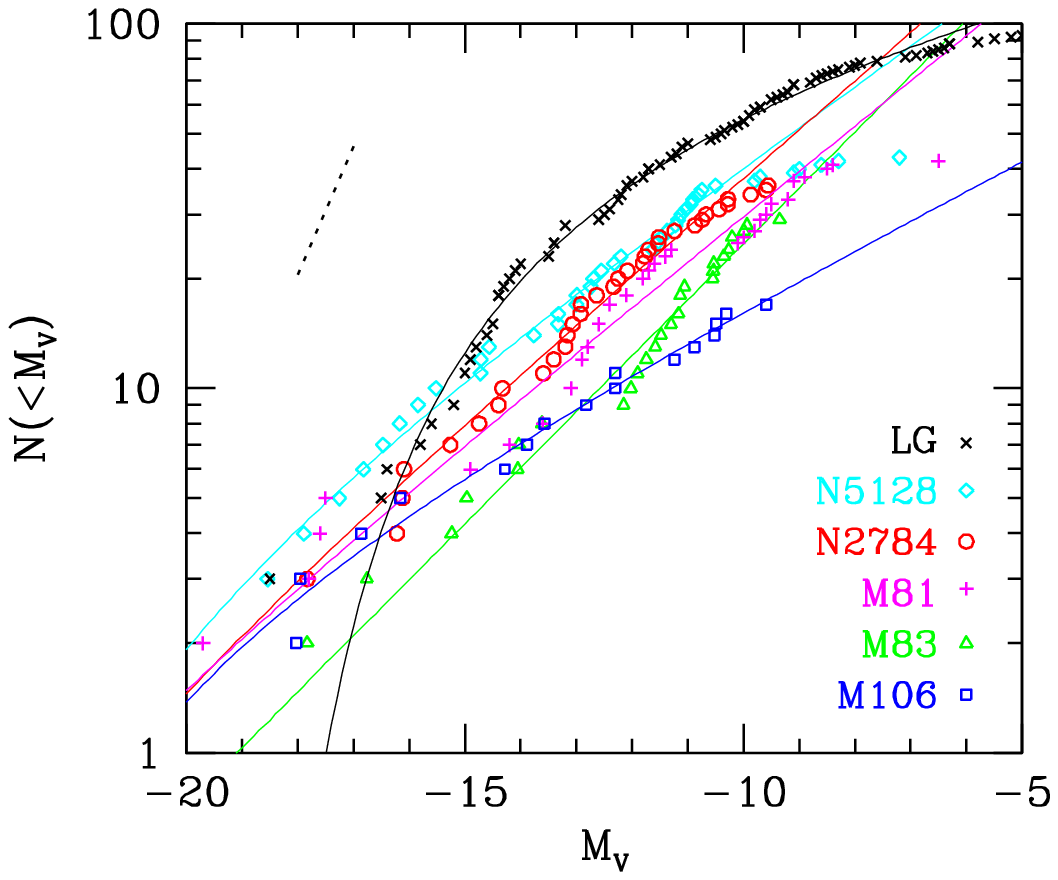}
\caption{Cumulative luminosity function of galaxies in several galaxy groups. 
The crosses, diamonds, circles, pluses, triangles, and boxes represent the galaxies 
in the Local Group \citep{mcc12,mar13a,mar13b}, NGC 5128 group \citep{tul15,crn16}, 
  NGC 2784 group (this study), M81 group \citep{chi09}, M83 group \citep{mul15}, and M106  group \citep{kim11}, respectively.
The solid lines indicate the best fitting cumulative Schechter function on each data.
The dotted line represents the faint-end slope ($\alpha=-1.8$) of halo mass expected from the $\Lambda$CDM model \citep{tre02}.  
\label{fig-clf}}
\end{figure}
\clearpage

\begin{figure}
\epsscale{0.8}
\plotone{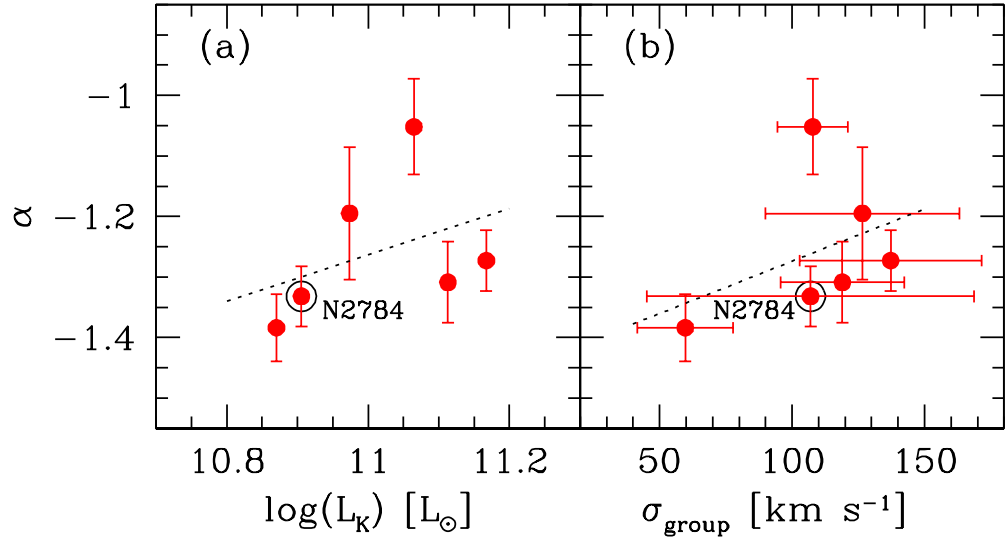}
\caption{Luminosity function slopes of the dwarf galaxies in the galaxy groups
as a function of (a) the total $K$-band luminosity of the groups ($L_K$)
and (b) the velocity dispersion of the member galaxies in each group ($\sigma_{group}$).
The filled circles with the large open circle indicate the measurement of the NGC 2784 group obtained in this study and the other filled circles are those of other groups obtained from the literature data.
The dotted lines represent the linear least-square fits to the data.
\label{fig-alpha}}
\end{figure}
\clearpage



\begin{thebibliography}{}

\bibitem[APASS (2015)]{apass} APASS, 2015. The AAVSO Photometric All-Sky Survey: Data Release 9 https://www.aavso.org/apass/

\bibitem[Bechtol et al.(2015)]{bec15} Bechtol, K., Drlica-Wagner, A., Balbinot, E., et al.\ 2015, \apj, 807, 50 


\bibitem[Behroozi et al.(2010)]{beh10} Behroozi, P.~S., Conroy, C., \& Wechsler, R.~H.\ 2010, \apj, 717, 379 

\bibitem[Benson et al.(2002)]{ben02} Benson, A.~J., Lacey, C.~G., Baugh, C.~M., Cole, S., \& Frenk, C.~S.\ 2002, \mnras, 333, 156 

\bibitem[Bertin \& Arnouts (1996)]{ber96} Bertin, E. \& Arnouts, S. 1996: SExtractor: Software for source extraction, Astronomy \& Astrophysics Supplement 317, 393 

\bibitem[Bertin et al. (2002)]{ber02} Bertin, E. et al. 2002, The TERAPIX Pipeline, ASP Conference Series, Vol. 281, 2002 D.A. Bohlender, D. Durand, and T.H. Handley, eds., p. 228

\bibitem[Bertin et al. (2006)]{ber06}  Bertin, E. 2006: Automatic Astrometric and Photometric Calibration with SCAMP, ASP Conference Series, Vol. 351, 2006, C. Gabriel, C. Arviset, D. Ponz, and E. Solano, eds., p. 112

\bibitem[Binggeli \& Jerjen(1998)]{bin98} Binggeli, B., \& Jerjen, H.\ 1998, \aap, 333, 17 

\bibitem[Blanton et al.(2005)]{bla05} Blanton, M.~R., Lupton, R.~H., Schlegel, D.~J., et al.\ 2005, \apj, 631, 208 


\bibitem[Chiboucas et al.(2009)]{chi09} Chiboucas, K., 
Karachentsev, I.~D., \& Tully, R.~B.\ 2009, \aj, 137, 3009 

\bibitem[Cooper et al.(2010)]{coo10} Cooper, A.~P., Cole, S., Frenk, C.~S., et al.\ 2010, \mnras, 406, 744 

\bibitem[Crnojevi{\'c} et al.(2016)]{crn16} Crnojevi{\'c}, 
D., Sand, D.~J., Spekkens, K., et al.\ 2016, arXiv:1512.05366 

\bibitem[Doyle et al.(2005)]{doy05} Doyle, M.~T., Drinkwater, M.~J., Rohde, D.~J., et al.\ 2005, \mnras, 361, 34

\bibitem[Drlica-Wagner et al.(2015)]{drl15} Drlica-Wagner, A., Bechtol, K., Rykoff, E.~S., et al.\ 2015, \apj, 813, 109 


\bibitem[Ferguson \& Sandage(1988)]{fer88} Ferguson, H.~C., \& Sandage, A.\ 1988, \aj, 96, 1520 

\bibitem[Ferrarese et al.(2006)]{fer06} Ferrarese, L., C{\^o}t{\'e}, P., Jord{\'a}n, A., et al.\ 2006, \apjs, 164, 334 

\bibitem[Ferrarese et al.(2016)]{fer16} Ferrarese, L., C{\^o}t{\'e}, P., S{\'a}nchez-Janssen, R., et al.\ 2016, \apj, 824, 10 

\bibitem[Fisher \& Tully(1981)]{fis81} Fisher, J.~R., \& Tully, R.~B.\ 1981, \apjs, 47, 139

\bibitem[Fouque et al.(1992)]{fou92} Fouque, P., Gourgoulhon, E., Chamaraux, P., \& Paturel, G.\ 1992, \aaps, 93, 211 

\bibitem[Garcia(1993)]{gar93} Garcia, A.~M.\ 1993, \aaps, 100, 47 

\bibitem[He et al.(2016)]{he16} He, M.~Y., Moon, D.-S., Neilson, H., et al.\ 2016, Journal of Korean Astronomical Society, 49, 209 


\bibitem[Jerjen et al.(2000)]{jer00} Jerjen, H., Binggeli, 
B., \& Freeman, K.~C.\ 2000, \aj, 119, 593 


\bibitem[Lee et al.(2016)]{lee16} Lee, Y., Rey, S.-C., Hilker, M., Sheen, Y.-K., \& Yi, S.~K.\ 2016, \apj, 822, 92 

\bibitem[Lisker et al.(2008)]{lis08} Lisker, T., Grebel, E.~K., \& Binggeli, B.\ 2008, \aj, 135, 380 

\bibitem[Lisker et al.(2013)]{lis13} Lisker, T., Weinmann, S.~M., Janz, J., \& Meyer, H.~T.\ 2013, \mnras, 432, 1162 

\bibitem[Lupton (2009)]{lup05} Lupton, R. 2005, Transformations between SDSS magnitudes and other systems https://www.sdss3.org/dr10/algorithms/sdssUBVRITransform.php/

\bibitem[Karachentsev et al.(2008)]{kar08} Karachentsev, I.~D., Makarov, D.~I., Karachentseva, V.~E., \& Melnik, O.~V.\ 2008, Astronomy Letters, 34, 832

\bibitem[Karachentsev et al.(2013)]{kar13} Karachentsev, 
I.~D., Makarov, D.~I., \& Kaisina, E.~I.\ 2013, \aj, 145, 101 

\bibitem[Kim et al.(2011)]{kim11} Kim, E., Kim, M., Hwang, 
N., et al.\ 2011, \mnras, 412, 1881 

\bibitem[Kim et al.(2016)]{kim16} Kim, S.-L., et al.\ 2016, JKAS, 49, 37 

\bibitem[Klypin et al.(1999)]{kly99} Klypin, A., Kravtsov, 
A.~V., Valenzuela, O., \& Prada, F.\ 1999, \apj, 522, 82 

\bibitem[Koposov et al.(2008)]{kop08} Koposov, S., Belokurov, V., Evans, N.~W., et al.\ 2008, \apj, 686, 279-291 

\bibitem[Makarov \& Karachentsev (2011)]{mak11} Makarov, D., \& Karachentsev, I.\ 2011, \mnras, 412, 2498 

\bibitem[Markwardt(2009)]{mar09} Markwardt, C.~B.\ 2009, ASPC, Astronomical Data Analysis Software and Systems XVIII, 411, 251

\bibitem[Martin et al.(2013a)]{mar13a} Martin, N.~F., Schlafly, E.~F., Slater, C.~T., et al.\ 2013a, \apjl, 779, L10 

\bibitem[Martin et al.(2013b)]{mar13b} Martin, N.~F., Slater, C.~T., Schlafly, E.~F., et al.\ 2013b, \apj, 772, 15 

\bibitem[McConnachie (2012)]{mcc12} McConnachie, A.~W.\ 2012, 
\aj, 144, 4  

\bibitem[McConnachie et al.(2009)]{mcc09} McConnachie, A.~W., Irwin, M.~J., Ibata, R.~A., et al.\ 2009, \nat, 461, 66 

\bibitem[Merritt et al.(2014)]{mer14} Merritt, A., van 
Dokkum, P., \& Abraham, R.\ 2014, \apjl, 787, L37 

\bibitem[Moon et al.(2016)]{moo16} Moon, D.-S., Kim, S. C., Lee, J.-J., et al.\ 2016, 
Proc. SPIE, 9906, 99064I

\bibitem[Moore et al.(1999)]{moo99} Moore, B., Ghigna, S., 
Governato, F., et al.\ 1999, \apjl, 524, L19 


\bibitem[M{\"u}ller et al.(2015)]{mul15} M{\"u}ller, O., 
Jerjen, H., \& Binggeli, B.\ 2015, \aap, 583, A79 

\bibitem[M{\"u}ller et al.(2017)]{mul17} M{\"u}ller, O., Jerjen, H., \& Binggeli, B.\ 2017, \aap, 597, A7 

\bibitem[Mu{\~n}oz et al.(2015)]{mun15} Mu{\~n}oz, R.~P., Eigenthaler, P., Puzia, T.~H., et al.\ 2015, \apjl, 813, L15 

\bibitem[Pak et al.(2014)]{pak14} Pak, M., Rey, S.-C., Lisker, T., et al.\ 2014, \mnras, 445, 630 

\bibitem[Popesso et al.(2006)]{pop06} Popesso, P., Biviano, A., B{\"o}hringer, H., \& Romaniello, M.\ 2006, \aap, 445, 29 

\bibitem[Presotto et al.(2012)]{pre12} Presotto, V., Iovino, A., Scodeggio, M., et al.\ 2012, \aap, 539, A55 


\bibitem[Roberts et al.(2015)]{rob15} Roberts, I.~D., Parker, L.~C., Joshi, G.~D., \& Evans, F.~A.\ 2015, \mnras, 448, L1 

\bibitem[Schechter(1976)]{sch76} Schechter, P.\ 1976, \apj, 203, 297 

\bibitem[Schlafly \& Finkbeiner(2011)]{sch11} Schlafly, E.~F., \& Finkbeiner, D.~P.\ 2011, \apj, 737, 103 

\bibitem[Smith et al.(2012)]{smi12} Smith, R.~J., Lucey, J.~R., Price, J., Hudson, M.~J., \& Phillipps, S.\ 2012, \mnras, 419, 3167 

\bibitem[Sprinthall (2011)]{spr11} Sprinthall, R. C. 2011, Basic Statistical Analysis (9th Edition), Pearson Education

\bibitem[Tolstoy et al.(2009)]{tol09} Tolstoy, E., Hill, V., \& Tosi, M.\ 2009, \araa, 47, 371 

\bibitem[Tonry et al.(2001)]{ton01} Tonry, J.~L., Dressler, 
A., Blakeslee, J.~P., et al.\ 2001, \apj, 546, 681 

\bibitem[Trentham \& Tully(2002)]{tre02} Trentham, N., \& Tully, R.~B.\ 2002, \mnras, 335, 712 

\bibitem[Trentham et al.(2005)]{tre05} Trentham, N., Sampson, L., \& Banerji, M.\ 2005, \mnras, 357, 783 

\bibitem[Trentham \& Tully(2009)]{tre09} Trentham, N., \& Tully, R.~B.\ 2009, \mnras, 398, 722 

\bibitem[Tully et al.(2015)]{tul15} Tully, R.~B., Libeskind, 
N.~I., Karachentsev, I.~D., et al.\ 2015, \apjl, 802, L25 

\bibitem[Weinmann et al.(2011)]{wei11} Weinmann, S.~M., Lisker, T., Guo, Q., Meyer, H.~T., \& Janz, J.\ 2011, \mnras, 416, 1197 

\bibitem[Wetzel et al.(2012)]{wet12} Wetzel, A.~R., Tinker, J.~L., \& Conroy, C.\ 2012, \mnras, 424, 232 

\bibitem[Wetzel et al.(2013)]{wet13} Wetzel, A.~R., Tinker, J.~L., Conroy, C., \& van den Bosch, F.~C.\ 2013, \mnras, 432, 336 
 
\end{thebibliography}
\end{document}